\documentclass[12pt]{article}
\newcommand{\df}{\ {\overset {\rm def} =}\ }
\newcommand{\dr}[2]{\frac {{\rm d} {#1}} {{\rm d} {#2}}}
\newcommand{\pdr}[2]{\frac {\partial {#1}} {\partial {#2}}}
\newcommand{\dril}[2]{{{\rm d} {#1}} / {{\rm d} {#2}}}

\newcommand{\llim}[1] {\ {\underset {#1} {\longrightarrow}}\ }

 \usepackage[dvips]{graphicx}
 \usepackage{amsmath}
 \usepackage{amsfonts}
 \usepackage{amssymb}

\begin{document}

\title{Avoidance of singularities in spherically symmetric charged dust}

\author{Andrzej Krasi\'nski and Krzysztof Bolejko\thanks{This research was
supported by the Polish Ministry of Education and Science grant no 1 P03B 075
29.}}

\date {}

\maketitle

\centerline{N. Copernicus Astronomical Center, Polish Academy of Sciences}

\centerline{Bartycka 18, 00 716 Warszawa, Poland}

\centerline{email: akr@camk.edu.pl}

\begin{abstract}
In spherically symmetric charged dust, just like in neutral dust, two kinds of
singularity may be present: the Big Bang/Crunch (BB/BC) singularity, and shell
crossings. Quite unlike in neutral dust, the BB/BC singularity may be avoided.
When the charge density $\rho_e$ and the mass-energy density $\epsilon$ obey
$\left|\rho_e\right| < D_e \df \sqrt{G} \epsilon / c^2$, the conditions that
allow the model to avoid the BB/BC singularity necessarily lead to shell
crossings. However, when $\left|\rho_e\right| \to D_e$ at the center of symmetry
while $\left|\rho_e\right| < D_e$ elsewhere, both kinds of singularity may be
avoided for a sufficiently long period that a body of charged dust may go
through the tunnel between the singularities in the maximally extended Reissner
-- Nordstr\"{o}m spacetime, to emerge into another asymptotically flat region.
An explicit example of such a configuration is presented and discussed. It does
not contradict any astrophysical constraints.
\end{abstract}

\section{Spherically symmetric dust in electromagnetic field}\label{emdust}

For the spherically symmetric spacetimes in the comoving coordinates, the metric
can be put in the form
\begin{equation}\label{1.1}
{\rm d} s^2 = {\rm e}^{C(t,r)} {\rm d} t^2 - {\rm e}^{A(t, r)} {\rm d} r^2 -
R^2(t,r)\left[{\rm d} \vartheta^2 + \sin^2(\vartheta) {\rm d} \varphi^2\right].
\end{equation}
The case $R,_r = 0$ requires separate consideration. In this case, a solution of
the Einstein equations for electrically neutral dust was found by Datt
\cite{Datt1938} and then derived and interpreted by Ruban \cite{Ruba1968,
Ruba1969}; the corresponding Einstein--Maxwell equations were systematically
derived and solved by Ruban \cite{Ruba1972, Kras1997, PlKr2006}.

We will discuss the solutions of the Einstein--Maxwell equations in the generic
case $R,_r \neq 0$. Most of the material of Sections 2 to 7 has been known for a
long time, but a few points were overlooked by some of the earlier authors. That
material is reviewed here in order to introduce the framework and notation, and
to fill in the few little gaps. New results are reported in Sections 8 and
following.

The following problem is considered. In a spherically symmetric charged dust,
just like in neutral dust, two kinds of singularity may be present: the Big
Bang/Crunch (BB/BC), and shell crossings. Quite unlike in neutral dust, the
BB/BC singularity may be avoided. However, when the charge density $\rho_e$ and
the mass-energy density $\epsilon$ obey $\left|\rho_e\right| < D_e \df \sqrt{G}
\epsilon / c^2$, the conditions that allow the model to avoid the BB/BC
singularity necessarily lead to shell crossings. (The repulsive term that
prevents the BB/BC singularity in this case is a purely relativistic effect.)
There exist initial conditions that allow us to avoid both singularities -- when
either $\left|\rho_e\right| > D_e$ everywhere, or $\left|\rho_e\right| \to D_e$
at the center of symmetry while $\left|\rho_e\right| < D_e$ elsewhere. The first
case was considered by Ori \cite{Oriun}, but remained unpublished. In the second
case, both kinds of singularity may be avoided for a sufficiently long period
that a body of charged dust goes through the tunnel between the singularities in
the maximally extended Reissner -- Nordstr\"{o}m spacetime, to emerge into
another asymptotically flat region. An explicit example of such a configuration
is presented and discussed. It has $E < 0$, i.e. it belongs to the recollapsing
class that could possibly be pulsating for ever. However, shell crossings
prevent it from existing for longer than one full cycle of collapse and
expansion. As shown in the last section, the parameters of the object do not
violate any conventional astrophysical limitations.

It is generally expected that shell crossings are just an artefact of the
assumption that the spatial gradient of pressure is zero. In a more general
solution, they would be replaced by acoustic waves of a high but finite density
that become singularities in the limit of spatially homogeneous pressure (in
particular, of zero pressure). However, no exact solutions of this kind are
available, and the singularities in such configurations have so far been
considered only via existence theorems applied to Einstein's equations with
various sources, such as the Einstein -- Maxwell -- Vlasov equations treated in
Ref. \cite{Noun2005}. Typically, the existence theorems predict the properties
of solutions over a rather limited time-interval. Lacking any more general exact
solution, the charged dust models are the best device that is available for
investigating long-term evolution of charged bodies.

\section{Spherically symmetric dust models with $R,_r \neq 0$}\label{Eequs}

\setcounter{equation}{0}

With spherical symmetry, in the coordinates of (\ref{1.1}), the electromagnetic
tensor can have at most two nonzero components, $F_{tr}$ (the electric field)
and $F_{\vartheta \varphi}$ (the magnetic field). The $F_{\vartheta \varphi}
\neq 0$ is due to a distribution of magnetic monopoles. Since they are known
(from experiment) not to exist, and since their inclusion does not lead to any
significant generalisation\footnote{With magnetic monopoles taken into account,
the electric charge $Q$ has to be replaced in all formulae by $\sqrt{{Q_e}^2 +
{Q_m}^2}$, where $Q_e$ is the electric charge and $Q_m$ is the magnetic
charge.}, we will assume the magnetic monopole charge to be zero everywhere.

The Einstein--Maxwell equations are in this case
\begin{eqnarray}
R_{\alpha \beta} - \frac 1 2 g_{\alpha \beta} R + \Lambda g_{\alpha \beta} &=&
\frac {8\pi G} {c^4} \epsilon u_{\alpha} u_{\beta} \nonumber \\
     &+& \frac {2G} {c^4} \left({F_{\alpha}}^{\mu} F_{\mu \beta} + \frac 1 4
g_{\alpha \beta} F_{\mu \nu} F^{\mu \nu}\right), \label{2.1} \\
{F^{\mu \nu}};_{\nu} &=& (4\pi / c) \rho_e u^{\mu}, \label{2.2} \\
F_{[\mu \nu, \rho]} &=& 0, \label{2.3}
\end{eqnarray}
where $\epsilon$ is the mass density of dust, $u^{\alpha}$ is the velocity field
of dust, and $\rho_e$ is the density of the electric charge.

From (\ref{2.2}) we find
\begin{eqnarray}
F^{01} &=& Q(r) {\rm e}^{- (A + C)/2} / R^2, \label{2.4} \\
Q,_r &=& (4\pi / c) \rho_e {\rm e}^{A/2} R^2, \label{2.5}
\end{eqnarray}
where $Q(r)$ is an arbitrary function; it is the electric charge within the
$r$-surface as (\ref{2.5}) shows. With the assumed $F_{23} = 0$, eq. (\ref{2.3})
is fulfilled identically.

Now the coordinate components of the Einstein equations with the metric
(\ref{1.1}) and the electromagnetic tensor (\ref{2.4}) -- (\ref{2.5}) become
\begin{eqnarray}
G_{00} &=& \frac {8\pi G} {c^4} \epsilon {\rm e}^C + \frac G {c^4}\ \frac
{Q^2} {R^4}\ {\rm e}^C - \Lambda {\rm e}^C, \label{2.6} \\
G_{01} &=& 0, \label{2.7} \\
G_{11} &=& - \frac G {c^4}\ \frac {Q^2} {R^4}\ {\rm e}^A + \Lambda
{\rm e}^A, \label{2.8} \\
G_{22} &\equiv& G_{33} / \sin^2(\vartheta) = \frac G {c^4}\ \frac {Q^2} {R^2} +
\Lambda R^2. \label{2.9}
\end{eqnarray}
For $G_{\alpha \beta}$ we find
\begin{eqnarray}
G_{00} &=& \frac {{\rm e}^C} {R^2} + \frac {{R,_t}^2} {R^2} + \frac
{A,_t R,_t} R \nonumber \\
     &\ &\ \ \ \ \ + {\rm e}^{C - A} \left(- \frac {{R,_r}^2} {R^2} - 2
\frac {R,_{r r}} R + \frac {A,_r R,_r} R\right),\ \ \ \ \ \ \label{2.10} \\
G_{01} &=& - 2 \frac {R,_{t r}} R + \frac {A,_t R,_r} R + \frac {R,_t C,_r} R,
\label {2.11} \\
G_{11} &=& - \frac {{\rm e}^A} {R^2} + \frac {{R,_r}^2} {R^2} +
\frac {C,_r R,_r} R \nonumber \\
     &\ &\ \ \ \ \ + {\rm e}^{A - C} \left(- \frac {{R,_t}^2} {R^2} - 2 \frac
{R,_{t t}} R + \frac {C,_t R,_t} R\right), \label{2.12} \\
G_{22} &=& \frac 1 4 {\rm e}^{- C} \left(- 4RR,_{t t} + 2R C,_t R,_t - 2RA,_t
R,_t\right. \nonumber \\
     &\ &\ \ \ \ \ \left.- R^2 {A,_t}^2 - 2R^2 A,_{t t} + R^2 C,_t A,_t\right)
\nonumber \\
\ \ \ &+& \frac 1 4 {\rm e}^{- A} \left(4RR,_{r r} + 2RC,_r R,_r - 2RA,_r R,_r
\right. \nonumber \\
     &\ &\ \ \ \ \ \left.+ R^2 {C,_r}^2 + 2R^2 C,_{r r} - R^2 C,_r A,_r\right).\
\ \ \ \ \ \label{2.13}
\end{eqnarray}
Equations (\ref{2.1}) -- (\ref{2.3}) imply the following:
\begin{eqnarray}
\left(\epsilon u^{\beta}\right);_{\beta} &=& 0, \label{2.14} \\
\epsilon {u^{\alpha}};_{\beta} u^{\beta} &=& - \frac 1 c \ \rho_e
{F^{\alpha}}_{\beta} u^{\beta}. \label{2.15}
\end{eqnarray}
Equations (\ref{2.14}) (the conservation of mass) and (\ref{2.15}) (the Lorentz
force acting on charges in motion and pushing them off geodesic trajectories)
are quite general and independent of any symmetry properties of spacetime.
Applying (\ref{2.15}) to our metric (\ref{1.1}) we get
\begin{equation}\label{2.16}
\epsilon C,_r = \frac 1 {2 \pi} \frac {QQ,_r} {R^4}.
\end{equation}
Applying (\ref{2.14}) to our metric (\ref{1.1}) we then get
\begin{equation}\label{2.17}
\frac {\kappa} 2 \epsilon R^2 {\rm e}^{A/2} = \frac G {c^4}\ N,_r,
\end{equation}
where $N,_r$ is an arbitrary function of integration. We see that $N$ so defined
is the energy equivalent to the sum of rest masses within the $r$-surface. From
(\ref{2.5}) and (\ref{2.17}) we now see that the ratio $Q,_r/N,_r = \rho_e/(c
\epsilon)$ is time-independent.

Using now (\ref{2.16}) and (\ref{2.11}) we obtain from the equation $G_{01} = 0$
\begin{equation}\label{2.18}
2 {\rm e}^{- A/2} R,_{t r} - {\rm e}^{- A/2} A,_t R,_r = \frac 2 c\ \frac {R,_t}
{R^2}\ \left(\frac {\rho_e Q} {\epsilon}\right).
\end{equation}
(It is here that the case $R,_r = 0$ has to be set aside for separate
investigation. With $R,_r = 0$, the equation $G_{01} = 0$ reduces to $R,_t C,_r
= 0$ and cannot be used to determine $R,_r$ as we do below.) Since the
expression in parentheses is independent of $t$, this can be integrated with the
result
\begin{equation}\label{2.19}
{\rm e}^{-A/2}R,_r = \Gamma(r) - \frac {\rho_e Q} {c \epsilon R},
\end{equation}
where $\Gamma(r)$ is an arbitrary function of integration. Using (\ref{2.5}) and
(\ref{2.17}) we now find
\begin{equation}\label{2.20}
\frac {\rho_e Q} {c \epsilon} \equiv \frac {QQ,_r} {N,_r} = QQ,_N.
\end{equation}
With this, (\ref{2.19}) becomes
\begin{equation}\label{2.21}
{\rm e}^{-A/2}R,_r = \Gamma(r) - \frac {QQ,_N} R,
\end{equation}
and now (\ref{2.16}) becomes
\begin{equation}\label{2.22}
C,_r = 2 \frac {{\rm e}^{A/2}} {R^2} QQ,_N.
\end{equation}
Using (\ref{2.21}) and (\ref{2.22}) to eliminate $R,_r$ and $C,_r$ from
(\ref{2.12}) and integrating the $G_{11}$ equation we get
\begin{equation}\label{2.23}
{\rm e}^{- C}{R,_t}^2 = \Gamma^2 - 1 + \frac {2M(r)} R + \frac
{Q^2\left({Q,_N}^2 - G/c^4\right)} {R^2} - \frac 1 3 \Lambda R^2,
\end{equation}
where $M(r)$ is an arbitrary function. Comparing this with the Newtonian
equation of motion we see that $(\Gamma^2 - 1)/2$ plays here the role of the
energy function $E(r)$.

The function $M(r)$ is the {\it effective mass} that, together with $Q$, drives
the evolution. However, as we will see, $M(r)$ is a combination of mass and
charge that {\it need not be positive}. In order to see this, we compare
(\ref{2.23}) with the Newtonian limit, assuming $\Lambda = 0$. Let $[x]$ denote
the physical dimension (unit) of $x$. The dimensions of the quantities appearing
in (\ref{2.23}) are: $[R] = [{\rm length}]$, $[{\rm e}^{C/2} {\rm d} t] \equiv
[{\rm d} s] = [c] \times [{\rm time}]$, $[M] = [G] \times [{\rm mass}] / [c^2]$,
$[N] = [{\rm mass}] \times [c^2]$, $[Q] = [{\rm charge}] \equiv [\sqrt{\rm
mass}] \times [{\rm length}^{3/2}] / [{\rm time}]$. The function $\left(\Gamma^2
- 1\right)$ is dimensionless, but, for consistency, must be assumed to have the
form $2 {\cal E} / c^2$, where $[{\cal E}] = [{\rm velocity}^2]$. In order to
find the Newtonian limit of (\ref{2.23}), we multiply it by $c^2$ and let $c \to
\infty$. Denoting the Newtonian time by $\tau$ we obtain
\begin{equation}\label{2.24}
{R,_{\tau}}^2 = 2 {\cal E} + 2 Gm(r) / R.
\end{equation}
But the Newtonian equation of motion of spherically symmetric charged dust is
\begin{equation}\label{2.25}
{r,_{\tau}}^2 = 2 {\cal E} (r) + 2 \left[G {\cal M} (r) - \frac {\rho_e(r)}
{\rho_{\mu}(r)}\ Q(r)\right] / r,
\end{equation}
where ${\cal M} (r)$ is the total mass within the sphere of radius $r$, $Q(r)$
is the total charge within the same sphere, $\rho_e$ is the charge density and
$\rho_{\mu}$ is the mass density. Thus, the $M(r)$ in (\ref{2.23}) corresponds
to the Newtonian $\left[G {\cal M} (r) - \left(\rho_e(r) / \rho_{\mu}(r)\right)
Q(r)\right]$, and, as announced, does not have to be positive.

To verify the $G_{22}$ equation, we have to find $C,_t$ from the $G_{11}$
equation and $A,_t$ from the $G_{01} = 0$ equation. Substituting these, then
finding $A,_r$ from (\ref{2.21}), using the $r$-derivative of (\ref{2.23}) to
eliminate $R,_tR,_{tr}$, and again using (\ref{2.21}) to eliminate ${\rm e}^{-
A}$, we obtain
\begin{equation}\label{2.26}
QQ,_N \left[- \frac G {c^4} \Gamma N,_r + \left(M + QQ,_N\Gamma\right),_r\right]
= 0.
\end{equation}
One solution of this is $Q,_N = 0$, i.e. a constant total charge. We will
mention this simpler case later. When $Q,_N \neq 0$,
\begin{equation}\label{2.27}
\frac G {c^4} \Gamma N,_r = \left(M + QQ,_N\Gamma\right),_r.
\end{equation}
The quantity
\begin{equation}\label{2.28}
{\cal M} \df M + QQ,_N\Gamma,
\end{equation}
an exact analogue of the Newtonian ${\cal M}(r)$ of (\ref{2.25}), will appear
again in Sec. \ref{chrnmatch}. There we will find that ${\cal M}$ is the active
gravitational mass. Thus, via (\ref{2.27}), $\Gamma$ determines by how much
${\cal M}$ increases when a unit of rest mass is added to the source, i.e.
$\Gamma$ is a measure of the gravitational mass defect/excess. Solutions with
$\Gamma = 0$ are known, this is the Datt -- Ruban class \cite{Datt1938} --
\cite{Ruba1972}. Negative $\Gamma$ can also occur (see subsection 18.20.2 of
Ref. \cite{PlKr2006}). However, the case $\Gamma > 0$ corresponds to the most
ordinary configuration.

The final equation to take into account is (\ref{2.6}), and it reproduces
(\ref{2.17}). Using (\ref{2.21}) to eliminate ${\rm e}^{A/2}$ from (\ref{2.17}),
we get $\epsilon$ in an equivalent form:
\begin{equation}\label{2.29}
\kappa \epsilon = \frac {2GN,_r} {c^4 R^2 R,_r}\ \left(\Gamma - \frac {QQ,_N}
R\right).
\end{equation}

Finally, the Einstein--Maxwell equations for charged dust reduced to a set that
defines the functions $A(t, r)$, $C(t, r)$ and $R(t, r)$ implicitly. The
solution is constructed as follows:

1. Choose the functions $N(r)$, $Q(r)$ and $\Gamma(r)$, and then solve
(\ref{2.27}) to find $M(r)$. In fact, since the coordinate $r$ is determined up
to the transformations $r' = f(r)$, where $f$ is an arbitrary function, we can
choose $N$, $Q$ or $\Gamma$ as the radial coordinate. In the example in Sec.
\ref{Anexample}, our radial coordinate will be $N(r)$.

2. Given these, express ${\rm e}^{A/2}$ through $R$ via (\ref{2.21}).

3. The set of equations (\ref{2.22}) -- (\ref{2.23}) then defines ${\rm e}^C$
and $R$. In solving (\ref{2.22}) numerically for $C(t, r)$, an initial condition
has to be assumed. Note that so far $C$ is not defined uniquely -- the
coordinate $t$ can be transformed by $t = g(t')$, and then $C$ transforms by $C
= C' - 2 \ln \left(g,_{t'}\right)$. Using this, we can require that at the
center of symmetry $C(t, r_c) = 0$, i.e. for the particle that remains all the
time at the center the proper time $s = t$ -- the time coordinate in spacetime.

If $Q,_N = 0$, then $Q = $ const, i.e. $\rho_e = 0$ from (\ref{2.5}). This case
is the neutral dust moving in the exterior electric field of a charge
concentrated at $R = 0$.

With vanishing charge, eq. (\ref{2.15}) reduces to the equation of a geodesic,
while eqs. (\ref{2.21}), (\ref{2.22}) and (\ref{2.23}) reduce to those defining
the the Lema\^{\i}tre --Tolman model \cite{Kras1997, PlKr2006}.

\section{Matching the charged dust metric to the Reissner--Nordstr\"{o}m
metric}\label{chrnmatch}

\setcounter{equation}{0}

We first transform the R--N metric with $\Lambda$ to the appropriate
coordinates. For the standard curvature coordinates we write $(\tau, R_{RN})$,
and we introduce the symbol
\begin{equation}\label{3.1}
h \df 1 - \frac {2m} {R_{RN}} + \frac {e^2} {{R_{RN}}^2} + \frac 1 3 \Lambda
{R_{RN}}^2.
\end{equation}
We demand that the new coordinates $(t, r)$ are still orthogonal, so that
$g_{tr} = 0$, i.e.
\begin{equation}\label{3.2}
h \tau,_t \tau,_r - \frac 1 h R_{RN,t} R_{RN,r} = 0.
\end{equation}
The function $C_{RN}(t, r)$ is
\begin{equation}\label{3.3}
{\rm e}^{C_{RN}} = h {\tau,_t}^2 - \frac 1 h {R_{RN,t}}^2.
\end{equation}
We now solve (\ref{3.2}) -- (\ref{3.3}) for $\tau,_r$ and calculate
\begin{equation}\label{3.4}
\left(g_{11}\right)_{RN} \equiv h{\tau,_r}^2 - \frac 1 h {R_{RN,r}}^2 = - \frac
{{\rm e}^C {R_{RN,r}}^2} {h {\rm e}^C + {R_{RN,t}}^2}.
\end{equation}

The component $\left(g_{00}\right)_{RN}$ is not fully determined at this point,
we will determine it later. Since $\tau$ is defined by the partial differential
equation (\ref{3.2}), it still involves an arbitrary function of one variable.

We wish to match the charged dust metric of Sec. 1 to the R--N solution given
above across a hypersurface $r = r_b$. This requires that the induced 3-metric
and the second fundamental form of this hypersurface are the same for both
spacetime metrics. Continuity of the 3-metric requires that
\begin{equation}\label{3.5}
{\rm e}^{C(t, r_b)} = {\rm e}^{C_{RN}(t, r_b)}, \qquad R(t, r_b) = R_{RN} (t,
r_b).
\end{equation}
The transformations that keep the metric diagonal are still allowed.
Transforming $t$ by $t' = \int {\rm e}^{\alpha(t)} {\rm d} t$, where $\alpha =
C_{RN}(t, r_b) - C(t, r_b)$ we fulfil the first of (\ref{3.5}), while $g'_{11}$
and $g'_{01}$ are not changed.

On the surface $r = r_b$, $R_{RN}(t, r_b)$ must be the same function of $t$ as
$R(t,r_b)$. Consequently, $R,_t(t, r_b)$ must be the same in both metrics, and
so $R_{RN}(t, r_b)$ must obey
\begin{eqnarray}\label{3.6}
{\rm e}^{- C_{RN}(t, r_b)}{R_{RN,t}}^2(t, r_b) &=& \Gamma^2(r_b) - 1 + \frac
{2M(r_b)} {R_{RN}(t, r_b)} - \frac {G Q^2(r_b)} {c^4 {R_{RN}}^2(t, r_b)}
\nonumber \\
     &\ &\ \ \ \ \ - \frac 1 3 \Lambda {R_{RN}}^2(t, r_b).
\end{eqnarray}
The unit vector normal to the hypersurface $r = r_b$ has components
\begin{equation}\label{3.7}
X^{\mu} = \left(0, {\rm e}^{- A/2}, 0, 0\right) \equiv \left(0,
\left.\left(\Gamma - \frac {QQ,_N} R\right)\right/ R,_r, 0, 0\right)
\end{equation}
for the interior metric, and, from (\ref{3.4}),
\begin{equation}\label{3.8}
X_{RN}^{\mu} = \left(0, \frac {\sqrt{h + {\rm e}^{- C_{RN}} {{R_{RN}},_t}^2}}
{R_{RN,r}}, 0, 0\right)
\end{equation}
for the R--N solution. From the continuity of the second fundamental form we
have $\left.R,_r X^r\right|_{r = r_b} = \left.R_{RN,r} X_{RN}^r\right|_{r =
r_b}$ and $\left.\left({\rm e}^C\right),_r X^r\right|_{r = r_b} =
\left.\left({\rm e}^{C_{RN}}\right),_r X_{RN}^r\right|_{r = r_b}$. The first
condition says
\begin{equation}\label{3.9}
\left(h + {\rm e}^{- C_{RN}} {R_{RN,t}}^2\right)_{r = r_b} = \left(\Gamma -
\frac {QQ,_N} R\right)^2_{r = r_b},
\end{equation}
which ensures the continuity of $g_{11} = - {\rm e}^A$ across $r = r_b$, even
though we have not required this. Substituting for ${R_{RN,t}}^2$ from
(\ref{3.6}) and for $h$ from (\ref{3.1}), then comparing the coefficients we
obtain
\begin{equation}\label{3.10}
e = \frac {\sqrt{G}} {c^2} Q(r_b), \qquad m = \left(M + QQ,_N \Gamma\right)_{r =
r_b}.
\end{equation}

The continuity of the second fundamental form imposes one last condition, on
$C,_r$. Using (\ref{2.22}), (\ref{2.21}) and (\ref{3.4}), the condition is
\begin{equation}\label{3.11}
\left.C_{RN,r}\ \frac {\Gamma - QQ,_N/R} {R_{RN,r}}\right|_{r = r_b} = 2\frac
{QQ,_N} {R^2}.
\end{equation}
We have no expression yet for $C_{RN,r}$, and we will find it from the field
equations now. We know that ${\rm e}^A$ is continuous at $r = r_b$, so we can
use (\ref{2.21}) for $A_{RN}(t, r_b)$. Substitute this in (\ref{2.11}), and take
the equation $G_{01} = 0$ at $r = r_b$. The result is
\begin{equation}\label{3.12}
C_{RN,r}(t, r_b) = \left.\frac {2R_{RN,r} QQ,_N} {R^2 \left(\Gamma -
QQ,_N/R\right)}\right|_{r = r_b},
\end{equation}
and it shows that (\ref{3.11}) is fulfilled.

Thus, (\ref{3.10}) are the only limitations imposed on the charged dust metric
by the matching conditions. This matching was first discussed by Vickers
\cite{Vick1973}. The second of (\ref{3.10}) reveals the connection between the
active gravitational mass (in this case $m$) and the effective mass (compare
(\ref{2.28})).

\section{Prevention of the Big Crunch singularity by electric
charge}\label{bcprevent}

\setcounter{equation}{0}

In the present section we will deal only with the Big Bang/Big Crunch
singularities, assuming $\Lambda = 0$. The shell crossings will be discussed
separately in Sec. \ref{rnshellcr}. In the following we shall denote $E(r) =
\left(\Gamma^2 - 1\right)/2$.

The presence or absence of a singularity is detected by investigating the roots
of the right-hand side of (\ref{2.23}), which, for this purpose, is more
conveniently written as
\begin{equation}\label{4.1}
{\rm e}^{- C} R^2 {R,_t}^2 = 2E(r)R^2 + 2M(r)R + Q^2\left({Q,_N}^2 -
G/c^4\right) \df W(R).
\end{equation}
At each root of $W(R)$, the sign of $R,_t$ changes, and evolution is possible
only in those regions where $W(R) \geq 0$. The following cases occur

\medskip

(a) When $E < 0$, $W(R)$ has roots only if
\begin{equation}\label{4.2}
M^2 \geq 2EQ^2\left({Q,_N}^2 - G/c^4\right).
\end{equation}
With no roots, $W(R)$ would be negative at all $R$, so (\ref{4.2}) is the
condition for the existence of a solution of (\ref{4.1}). The roots are
\begin{equation}\label{4.3}
R_{\pm} = - \frac M {2E} \pm \frac 1 {2E} \sqrt{M^2 - 2EQ^2\left({Q,_N}^2 -
G/c^4\right)},
\end{equation}
and $W(R) > 0$ between them. Nonsingular solutions will exist when both $R_{\pm}
> 0$ (with $R_{\pm} < 0$, no solution exists at all, and with $R_- R_+ < 0$,
$R = 0$ is in the allowed range.) This is equivalent to
\begin{equation}\label{4.4}
{Q,_N}^2 < G/c^4 \qquad {\rm and}\qquad M > 0.
\end{equation}
We will interpret this condition later on in this section.

If there is equality in (\ref{4.2}), then $W(R) < 0$ for all $R \neq R_- = R_+$,
and $W(R_{\pm}) = 0$. Then $R = R_{\pm}$ and the model is static. If, in
addition, $Q,_N(r_b) = 0$ (meaning $\rho_e(r_b) = 0$) and $\Gamma(r_b) = 0$,
then $E = -1/2$, and in this case the exterior R--N metric is the extreme one,
with $e^2 = m^2$.

With (\ref{4.4}) fulfilled, $R$ oscillates between a nonzero minimum and a
maximum.

(b) When $E = 0$, singularity is avoided if and only if $M > 0$ and ${Q,_N}^2 <
G/c^4$. Collapse is then halted and reversed once and for all.

(c) When $E > 0$, $W(R) > 0$ either everywhere (if there are no roots) or beyond
the roots. There will be no roots when $M^2 < 2EQ^2\left({Q,_N}^2 - \right.$
$\left.G/c^4\right)$, in which case $W(R) > 0$ for all $R$ including $R = 0$,
and the model can run into the singularity. Thus (\ref{4.2}) is here one of the
necessary conditions for the existence of nonsingular solutions. With
(\ref{4.2}) fulfilled, $W(R)$ has two roots, and at least one of them has to be
positive if singularity is to be avoided. With $M > 0$, we have $R_- < 0$ always
and $R_+ > 0$ if and only if ${Q,_N}^2 < G/c^4$. With $M < 0$, $R_+ > 0$ and
$R_- < R_+$ always, so nonsingular solutions exist with no further conditions,
provided $R > R_+$ initially. Collapse is then halted and reversed as in case
(b). The bounce with $M < 0$ is nonrelativistic, since it occurs also in
Newton's theory, under the same conditions.

\medskip

Now we will interpret the condition (\ref{4.4}). The inequality ${Q,_N}^2 < G /
c^4$ translates into $\left|\rho_e\right| < \sqrt{G} \epsilon / c$, which means
that the absolute value of the charge density is sufficiently {\it small} (but
nonzero) compared to the mass density. This kind of bounce is purely
relativistic, and it does not occur in the Newtonian limit: with $M > 0$, $R =
0$ is always in the allowed range of Newtonian solutions. The interpretation of
the relativistic bounce in the Newtonian terms is this: as seen from
(\ref{2.23}), the charges provide a correction to the effective mass $M$, so
that it becomes $\overline{M} = M + (1/2) Q^2 \left({Q,_N}^2 - G / c^4\right) /
R$. This correction is negative at small charge density (when ${Q,_N}^2 < G /
c^4$), so it weakens gravitation, thus helping the dust to bounce. However, at
large charge density (${Q,_N}^2 > G / c^4$), charges enhance the effective mass
and thus oppose bounce. (A similar phenomenon is encountered in the motion of
particles in the Reissner--Nordstr\"om spacetime, where an electric charge in
the source of the gravitational field creates effective antigravitation,
provided the charge is small enough compared to mass.) Nevertheless, in this
last case, the Newtonian electrostatic repulsion can prevail, provided $M < 0$
at the same time.

With $Q,_N = 0$ the BB/BC singularity is avoided in every case when a solution
exists. Thus, for neutral dust moving in an exterior electric field, the BB/BC
singularity never occurs. This was first found by Shikin \cite{Shik1972}. This
is a purely relativistic effect.

The above implies that with (\ref{4.4}) fulfilled a solution of (\ref{2.23}),
for which $R \neq 0$ initially, does not go down to $0$. However, if the charged
dust occupies a volume around the center of symmetry $R = 0$, then, at any time,
there are dust particles with all values of $R$, including $R = 0$. (We will
find in the next section the conditions to be obeyed in order that the center is
nonsingular.) Thus, the inner turning points given by (\ref{4.3}) will exist
arbitrarily close to the center. We will use this remark in Sec.
\ref{rnshellcr}.

If $EM < 0$ and ${Q,_N}^2 = G/c^4$, then (\ref{4.1}) has the time-independent
solution $R = - M/E$. In this case, the electrostatic repulsion just balances
the gravitational attraction and the whole configuration is static -- but
unstable. When $M > 0 > E$, the perturbation can only be toward smaller $R$, and
it will send the dust into collapse that will terminate at $R = 0$. When $M < 0
< E$, the perturbation can only be toward larger $R$, and it will send the dust
into infinite expansion. Another time-independent solution is $E = 0 = M$,
${Q,_N}^2 = G / c^4$; in this case $G \Gamma / c^4 = D =$ const and $N = D {\cal
M}$.

The surface of the charged sphere obeys the equation of radial motion of a
charged particle in the Reissner--Nordstr\"{o}m spacetime. For such a particle,
if the ratio of its charge $q$ to its mass $\mu$ obeys $(q / \mu)^2 < 1$, then
the reversal of fall to escape can occur only inside the inner R--N horizon, at
$R < r_- = m - \sqrt{m^2 - e^2}$. Thus, the surface of a collapsing sphere must
continue to collapse until it crosses the inner horizon, and can bounce at $R <
r_-$. Then, however, it cannot re-expand back into the same spacetime region
from which it collapsed, as this would require motion backward in time, as seen
from the compactified spacetime diagram of the maximally extended R--N solution
\cite{PlKr2006}. The surface would thus continue through the tunnel between the
singularities and re-expand into another copy of the asymptotically flat region.
This possibility is interesting, since it shows that the maximally extended R--N
spacetime, usually interpreted as an abstract geometric curiosity, may in this
case become an astrophysical reality.

The bounce at small charge density (${Q,_N}^2 < G / c^4$) is more interesting
physically, since the real Universe has no detectable net charge, so only small
localized charges could exist in it. We saw that {\it an arbitrarily small
uncompensated charge can prevent the BB/BC singularity}. Unfortunately, Ori
\cite{Ori1990, Ori1991} proved that if ${Q,_N}^2 < G / c^4$ holds throughout the
volume, then a shell crossing is unavoidable, and it will block the passage
through the tunnel. We will derive this result in Sec. \ref{rnshellcr}. Thus, a
nonsingular bounce through the RN tunnel is possible only if ${Q,_N}^2 > G /
c^4$ everywhere or if ${Q,_N}^2 \to G / c^4$ at the center, while ${Q,_N}^2 < G
/ c^4$ elsewhere.

\section{Charged dust in curvature and mass-curvature coordinates}\label{rncurv}

\setcounter{equation}{0}

It is instructive to transform the metric given by (\ref{1.1}) with (\ref{2.21})
-- (\ref{2.23}) to such coordinates in which the function $R(t, r)$ is the
radial coordinate. We note that $R,_r {\rm d} r = {\rm d} R - R,_t {\rm d}t$,
and we take $t$ to be a function of the new coordinates: $t = f(\tau, R)$. Thus
\begin{eqnarray}\label{5.1}
R,_r {\rm d}r &=& {\rm d} R - R,_t \left(f,_{\tau} {\rm d} \tau + f,_R {\rm d}
R\right), \nonumber \\
{\rm d} t &=& f,_{\tau} {\rm d} \tau + f,_R {\rm d} R,
\end{eqnarray}
and the new metric components, using (\ref{2.23}), are found to be
\begin{eqnarray}
g_{\tau \tau} &=& {\rm e}^C {f,_{\tau}}^2 \Delta / \left(\Gamma -
QQ,_N/R\right)^2, \label{5.2} \\
g_{\tau R} &=& \frac {{\rm e}^C f,_{\tau} f,_R \Delta + f,_{\tau} R,_t}
{\left(\Gamma - QQ,_N/R\right)^2}, \label{5.3} \\
g_{R R} &=& \frac {{\rm e}^C {f,_R}^2 \Delta - 1 + 2f,_R R,_t} {\left(\Gamma -
QQ,_N/R\right)^2}, \label{5.4} \\
\Delta &\df& 1 - \frac {2 {\cal M}} R + \frac {GQ^2} {c^4 R^2} + \frac 1 3
\Lambda R^2, \label{5.5}
\end{eqnarray}
where ${\cal M}$ is defined in (\ref{2.28}). Note, from (\ref{5.1}), that the
transformation is different for collapsing dust ($R,_t < 0$) and for expanding
dust ($R,_t > 0$). The transformation from $(\tau, R)$ to $(t, r)$, inverse to
(\ref{5.1}), is analogous to introducing, in the Reissner--Nordst\"{o}m region
$r > r_+ = m + \sqrt{m^2 - e^2}$, coordinates comoving with the congruence of
charged particles that are radially collapsing or expanding, respectively. The
extension is to the future or to the past, respectively.

This can now be specialised in two ways. One possibility is to choose the proper
curvature coordinates, in which $g_{\tau R} = 0$. This representation of the
Vickers metric has not found any application in the literature so far, but it is
instructive -- see Ref. \cite{PlKr2006}. The other possible specialisation of
(\ref{5.2}) -- (\ref{5.4}) is to choose the ${\cal M}$ defined in (\ref{2.28})
as the new $\tau$ coordinate. These {\bf mass-curvature coordinates} were first
introduced by Ori \cite{Ori1990}. The surfaces ${\cal M} =$ const are timelike,
so none of the coordinates is time and the metric in these coordinates cannot be
diagonal. Since ${\cal M}$, $Q$, $\Gamma$ and $N$ depend only on $r$ in the
original coordinates, we have $Q = Q({\cal M})$, $\Gamma = \Gamma({\cal M})$ and
$N = N({\cal M})$. The Jacobi matrices of the transformations $(t, r)
\leftrightarrow ({\cal M}, R)$ are
\begin{equation}\label{5.6}
\pdr {(t, r)} {({\cal M}, R)} = \left[\begin{array}{l l} f,_{\cal M}, & f,_R \\
r,_{\cal M}, & r,_R \end{array}\right], \qquad \pdr {({\cal M}, R)} {(t, r)} =
\left[\begin{array}{l l} 0 & {\cal M},_r \\
R,_t, & R,_r \end{array}\right].
\end{equation}
These matrices must be inverse to each other, hence
\begin{equation}\label{5.7}
f,_{\cal M} = - \frac {R,_r} {R,_t {\cal M},_r}, \qquad f,_R = 1/R,_t, \qquad
r,_{\cal M} = 1 / {\cal M},_r, \qquad r,_R = 0.
\end{equation}
In the coordinates $(x^0, x^1) \df$ $({\cal M}, R)$ the velocity field still has
only one contravariant component:
\begin{equation}\label{5.8}
u^R = \pm \sqrt{\Gamma^2 - 1 + \frac {2M} R + \frac {Q^2\left({Q,_N}^2 -
G/c^4\right)} {R^2} - \frac 1 3 \Lambda R^2}
\end{equation}
($+$ for expansion, $-$ for collapse). We define the auxiliary quantities
\begin{equation}\label{5.9}
u \df \Gamma - QQ,_N/R,
\end{equation}
\begin{equation}\label{5.10}
\left({\rm e}^{C / 2} / u\right) f,_{\cal M} \df F({\cal M}, R),
\end{equation}
and, using (\ref{5.5}), (\ref{5.8}) and (\ref{2.23}), we get in (\ref{5.2}) --
(\ref{5.4}):
\begin{equation}\label{5.11}
g_{{\cal M} {\cal M}} = F^2 \Delta,  \qquad g_{{\cal M} R} = F u / u^R, \qquad
g_{R R} = 1 / (u^R)^2,
\end{equation}
while eqs. (\ref{5.1}) simplify to ${\rm d} r = {\rm d} {\cal M} / {\cal M},_r$
and ${\rm d} t = \left(- R,_r {\rm d} {\cal M} / {\cal M},_r + {\rm d} R\right)
/ R,_t$. The function $F({\cal M}, R)$ is to be found from the field equations.

Using (\ref{5.7}), (\ref{2.4}) and (\ref{5.10}) we find that the only
nonvanishing components of the electromagnetic tensor in the $({\cal M}, R)$
coordinates are
\begin{equation}\label{5.12}
F^{{\cal M} R} = - F^{R {\cal M}} = \frac Q {FR^2}, \qquad F_{{\cal M} R} = -
F_{R {\cal M}} = - \frac {FQ} {R^2}.
\end{equation}
Further, using (\ref{2.4}), (\ref{2.5}), (\ref{2.17}), (\ref{2.21}),
(\ref{2.27}), (\ref{2.23}) and (\ref{5.8}) we find for the charge density and
energy-density
\begin{equation}\label{5.13}
\frac {4\pi \rho_e} c = - \frac {Q,_{\cal M}} {R^2 F u^R}, \qquad \kappa
\epsilon = - \frac 2 {\Gamma R^2 F u^R}.
\end{equation}

Recall that the $({\cal M}, R)$-coordinates cover only such a region where
$R,_t$ has a constant sign. As seen from (\ref{5.7}) and (\ref{5.10}), $F$
changes sign where $R,_t$ does, and so does $u^R = {\rm e}^{- C/2} R,_t$. Thus,
$Fu^R$ preserves its sign when collapse turns to expansion and vice versa. This
observation will be useful when calculating $F$ later.

Now $F({\cal M}, R)$ is the only unknown function. We obtain further
\begin{eqnarray}
u_{\cal M} &=& uF, \qquad u_R = 1/u^R, \label{5.14} \\
g^{{\cal M} {\cal M}} &=& - \frac 1 {\left(Fu^R\right)^2}, \qquad g^{{\cal M} R}
= \frac u {Fu^R}, \qquad g^{RR} = - \Delta.\ \ \ \ \  \label{5.15}
\end{eqnarray}

In the $({\cal M}, R)$ coordinates eq. (\ref{2.27}) reads
\begin{equation}\label{5.16}
(G \Gamma / c^4) N,_{\cal M} = 1.
\end{equation}
The function $F$ can be found from (\ref{2.15}). Using (\ref{2.20}), it is found
to be:
\begin{equation}\label{5.17}
F,_R = - \frac {{u^R},_{\cal M}} {u \left(u^R\right)^2}.
\end{equation}
Using (\ref{5.9}) and (\ref{5.16}) we transform (\ref{5.17}) to
\begin{equation}\label{5.18}
F,_R = - \frac 1 {\left(u^R({\cal M}, R)\right)^3} \left\{\Gamma,_{\cal M} +
\frac 1 {R \Gamma}\left[1 - \frac {c^4} G \left({Q,_N}^2 +
QQ,_{NN}\right)\right]\right\}.
\end{equation}
This coincides, except for notation, with Ori's (1990) result. As Ori
\cite{Ori1990} stressed, eqs. (\ref{5.11}), (\ref{5.8}) and (\ref{5.18})
determine the metric explicitly, in contrast to the representation by Vickers
used in Sec. \ref{Eequs}, where the Einstein--Maxwell equations were reduced to
a set of two differential equations. However, we lost the information about the
time-dependence of $R$. Points of the spacetime are now identified by the values
of ${\cal M}$ and $R$ -- by specifying the pair $({\cal M}, R)$ we say: this is
the point in which the shell containing the mass ${\cal M}$ has the radius $R$.
However, we have no means of saying, without recourse to the comoving
coordinates, how much coordinate time it has taken the shell to expand from the
minimal size $R = R_{\rm min}$ to the current $R$. In consequence of this, the
$({\cal M}, R)$ coordinates do not allow us to see whether the minimal size
(which is a singularity or a nonsingular bounce) was achieved by all shells
simultaneously with respect to the time coordinate $t$ or not. This information
is crucial for considering shell crossings, as we will see in Sec.
\ref{rnshellcr}.

With $\Lambda = 0$, the integral of (\ref{5.18}) is elementary, but requires
separate treatment of various subcases. The full list of results is given in
Ori's paper.

Now the Einstein--Maxwell equations are all fulfilled.

Note from (\ref{5.11}) and (\ref{5.13}) that the metric and the mass density are
insensitive to the sign of $u^R$. However, as explained in the remark after
(\ref{5.5}), $u^R > 0$ and $u^R < 0$ correspond to different maps with different
domains. Thus, integrating (\ref{5.18}) from $R_1$ to $R > R_1$ with $u^R > 0$,
we integrate forward in time, while calculating the same integral with $u^R < 0$
we integrate backward in time.

\section{Regularity conditions at the center}\label{regcons}

\setcounter{equation}{0}

Just as in the L--T model, the set $R = 0$ in charged dust consists of the Big
Bang/Crunch singularity (which we showed to be avoidable) and of the center of
symmetry, which may or may not be singular. We will now derive the conditions
for the absence of the central singularity. We assume no magnetic charges.

Let $r = r_c$ correspond to the center of symmetry. From (\ref{1.1}),
(\ref{2.21}) and (\ref{2.29}) we see that $N(r) = \int_{\cal V} \epsilon
\sqrt{-g} {\rm d}_3 x = 4 \pi \int_0^r \epsilon(r') R^2(t,r') {\rm d} r'$, where
${\cal V}$ is a sphere centered at $r = r_c$ in a $t =$ const space. Thus, if
$\epsilon$ has no singularity of the type of the Dirac delta at the center,
$N(r)$ must obey $N(r_c)= 0$. Similarly, eqs. (\ref{1.1}) and (\ref{2.5}) show
that if there is no delta-type singularity of $\rho_e$ at the center, then the
electric charge must obey $Q(r_c) = 0$. With both $\epsilon$ and $\rho_e$ being
nonsingular at $r_c$, and $\epsilon(t, r_c) > 0$, the ratio $\rho_e(r_c) /
\epsilon(r_c)$ is nonsingular, and (\ref{2.5}) with (\ref{2.17}) show that
$\lim_{r \to r_c} Q/N = \lim_{r \to r_c}Q,_N = \rho_e(r_c) / [c \epsilon(r_c)]$
is finite (possibly zero). Then, (\ref{5.16}) implies that $\Gamma(r_c) \neq 0$
and $\lim_{r \to r_c} N/{\cal M} = \lim_{r \to r_c} N,_{\cal M} = G\Gamma /
c^4$, and this, together with $N\left(r_c\right) = 0$, implies ${\cal
M}\left(r_c\right) = 0$. Then, from (\ref{2.28}), also $M\left(r_c\right) = 0$.

Since $R(t, r_c) = 0$ and $N(r_c) = 0$, we find from (\ref{2.29}):
\begin{equation}\label{6.1}
\lim_{r \to r_c} \frac {R^3} N = \lim_{r \to r_c} \frac {3R^2 R,_r} {N,_r} =
\lim_{r \to r_c} \frac 3 {4 \pi \epsilon}\ \left(\Gamma - \frac {QQ,_N}
R\right).
\end{equation}
This limit will be finite if $\lim_{r \to r_c} (QQ,_N/R) < \infty$. We already
know that $Q,_N(r_c) \df {\widetilde q_0} =$ const and thus $\lim_{N \to 0} Q/N
= {\widetilde q_0}$. Thus, $\lim_{r \to r_c} Q/{\cal M} = q_0$ and $\lim_{r \to
r_c} (QQ,_N/R) < \infty$ if $\lim_{r \to r_c} R/{\cal M}^{\gamma} = {\rm
const}$, where $\gamma < 1$. Then, (\ref{6.1}) imposes the further condition
that, in the neighbourhood of $r_c$
\begin{equation}\label{6.2}
\lim_{r \to r_c} R/{\cal M}^{1/3} = \beta(t).
\end{equation}
We assume $\beta(t) \neq 0$, since $\beta = 0$ implies, via (\ref{6.1}), one of
two unphysical situations: (I) $\lim_{r \to r_c} \epsilon = \infty$, i.e. a
permanent central singularity, or (II) $\lim_{r \to r_c} \left(\Gamma -
QQ,_N/R\right) = 0$, which leads to $\epsilon(r_c)$ being independent of time --
a pathological situation in an expanding or contracting configuration.

Since $R(t, r_c) = 0$ at all times, we have $R,_t(t, r_c) = 0$, and, from the
above, $\lim_{r \to r_c}$ $R,_t/{\cal M}^{1/3} = \beta,_t$. All other terms in
(\ref{2.23}) except $(\Gamma^2 - 1)$ vanish at $r = r_c$, and so we must have
\begin{equation}\label{6.3}
\lim_{r \to r_c} \Gamma^2(r) = 1 \Longrightarrow \lim_{r \to r_c} E(r) = 0.
\end{equation}
From the limiting behaviour of the functions in (\ref{2.23}) at $r \to r_c$, we
conclude that
\begin{equation}\label{6.4}
\lim_{r \to r_c} 2E / {\cal M}^{2/3} = \lim_{r \to r_c} \left(\Gamma^2(r) -
1\right) / {\cal M}^{2/3} = {\rm const},
\end{equation}
and this constant may be zero. Note that with $\Gamma < 0$ this means
$\Gamma(r_c) = -1$. A central point where $\Gamma < 0$ is the ``second center of
symmetry'', in those models that have it. It is the antipodal point, in the
spherical space, to the ordinary center. Having reached the antipodal point, we
have added as much mass to the space as it can contain, and adding new mass is
not possible. The condition $\Gamma(r_c) = -1$ must then be understood as
follows: the active gravitational mass increases when we take away the rest mass
from the object.

\section{Shell crossings in charged dust}\label{rnshellcr}

\setcounter{equation}{0}

As is seen from (\ref{5.10}) and (\ref{5.7}), $F = 0$ is equivalent to $R,_r =
0$, so $F = 0$ is a locus of shell crossing. Then, from (\ref{5.13}) we see that
$\Gamma Fu^R$ must be negative for the density of dust to be positive. Since
$u^R = \dril R s < 0$ during collapse, $\Gamma F$ must then be positive.

We will investigate the occurrence of shell crossings in a configuration that
avoids the Big Bang/Crunch singularities. Let us write the solution of
(\ref{5.18}) as follows
\begin{equation}\label{7.1}
F = I({\cal M}, R, R_1) + g({\cal M}),
\end{equation}
where
\begin{equation}\label{7.2}
I({\cal M}, R, R_1) \df - \int_{R_1}^R \frac 1 {\left(u^R({\cal M}, x)\right)^3}
\left\{\Gamma,_{\cal M} + \frac 1 {x \Gamma}\left[1 - \frac {c^4} G
\left({Q,_N}^2 + QQ,_{NN}\right)\right]\right\} {\rm d} x,
\end{equation}
$R_1$ is the initial value of $R$ and $g({\cal M})$ is an arbitrary function --
the value of $F({\cal M} ,R)$ at $R = R_1$. In following the collapse of the
central region forward in time, we eventually come close to the bounce point,
where $R < R_1$. The integral becomes unbounded as $R$ approaches the bounce
value (where $u^R \to 0$), and so, with $R < R_1$, the first term in (\ref{7.1})
is positive when the expression in curly braces is negative.

We saw in Sec. \ref{bcprevent} that there are dust particles with all values of
$R$, including $R = 0$. At the turning point $u^R \to 0$. The integrand is of
the form $\left[\left(ax^2 + bx + c\right)^{-3/2} \left(a_2 x^3 + a_3
x^2\right)\right]$, and the trinomial has real zeros, so the integral is
unbounded, which shows that $u^R = 0$ is a coordinate
singularity.\footnote{Incidentally, the set where $\Gamma = 0$ is also a
coordinate singularity. This set is the ``neck'', known from studies of the
Lema\^{\i}tre--Tolman model, see Ref. \cite{KrHe2004}.} From (\ref{7.2}) it is
seen that as we get near to $\{R = 0, {\cal M} = 0\}$, the coefficient $[1 / (x
\Gamma)]$ will become unbounded. We know from the regularity conditions that
$\lim_{r \to r_c} \Gamma^2 = 1$, $\lim_{r \to r_c} Q = 0$, $\lim_{r \to r_c}
Q,_N =$ const $< \infty$ and $\lim_{r \to r_c} Q/R = 0$. Thus, as long as
${Q,_N}^2 < G/c^4$, the term containing $1/(x \Gamma)$ will dominate in the
vicinity of $R = 0$ and will determine the sign of the infinity in $F$. Now it
turns out that the sign of $F$ will necessarily change to opposite during
collapse: If $\Gamma > 0$ and $g({\cal M}) > 0$, then $F > 0$ at $R = R_1$, but
$F \to - \infty$ as $R \to 0$. If $\Gamma < 0$, then $F < 0$ can be achieved at
$R = R_1$ by the choice $g({\cal M}) < 0$, but $F \to + \infty$ as $R \to 0$.
This means, there is necessarily a shell crossing somewhere at $R
> 0$ if ${Q,_N}^2 < G/c^4$ holds all the way down to ${\cal M} = 0$. This is the
theorem proven by Ori \cite{Ori1990, Ori1991}.

The infinity in $F$ could be avoided if the term in curly braces in (\ref{7.2})
were zero at the same $x$, at which $u^R = 0$. The zero of $u^R$ is given by
(\ref{4.3}), it is $R = R_+$. In that case $F$ is finite at $R = R_+$, and, by
(\ref{5.13}), $\epsilon$ becomes infinite, i.e. $R = R_+$ becomes a true
curvature singularity. Thus, also in this case, the charged dust cannot tunnel
through the Reissner--Nordstr\"{o}m throat.

The only situations in which both BB/BC and shell crossing singularities could
possibly be avoided are these:

1. When $\lim_{r \to r_c} {Q,_N}^2 = G/c^4$ -- then, because of $\lim_{r \to
r_c} Q/R = 0$, the term $\Gamma,_{\cal M}$ in (\ref{7.2}) has a chance to
outbalance the other one and secure the right sign of $F$ everywhere.

2. When ${Q,_N}^2 > G / c^4$, $E > 0$ and $M < 0$. Then, as seen from
(\ref{4.3}), both $R_{\pm} > 0$ and a nonsingular bounce at $R_+$, with no shell
crossings, is possible.

Ori (private communication) found an example of a fully nonsingular bounce with
${Q,_N}^2 > G / c^4$, but has never published this result. An example of a
nonsingular bounce with ${Q,_N}^2 < G / c^4$ will be given in Sec.
\ref{Anexample}.

\section{The evolution of charged dust in mass-curvature coordinates.}

\setcounter{equation}{0}

In the mass-curvature coordinates, with $\Lambda = 0$, it is easy to solve the
evolution equation $\dril R s = u^R$. By this opportunity, we can again
identify, by another method, all those solutions that avoid the BB/BC
singularity.

We take $\ell = +1$ for expansion and $\ell = -1$ for collapse,\footnote{We
refer here to the initial instant of evolution. In the nonsingular models,
collapse will reverse into expansion during the evolution.} we denote
\begin{equation}\label{8.1}
2E \df \Gamma^2 - 1, \qquad \Phi \df \frac {Q^2 \left({Q,_N}^2 - G / c^4\right)}
{2E},
\end{equation}
and obtain from (\ref{5.8}) with $\Lambda = 0$
\begin{equation}\label{8.2}
\dr R s = \ell \sqrt {2E + 2M / R + 2E \Phi / R^2}.
\end{equation}
The integral is different for each sign of $E$, and for $E \geq 0$ there are
separate subcases depending on the values of $\Phi$ and $M$; $\Phi < 0$
corresponds to ${Q,_N}^2 < G/c^4$. With $\Phi = 0$ the solutions are of the same
form as those in the L--T model, although, if $Q \neq 0$, the effective mass $M$
still contains a contribution from the charges. With $E \neq 0$ it is most
convenient to represent the solution $R(s)$ by parametric formulae
$\left\{R(\omega), s(\omega)\right\}$.

With $E > 0$ and $\Phi < M^2 / \left(4 E^2\right)$, the solution is
\begin{eqnarray}\label{8.3}
R &=& \ell_2 \sqrt{\frac {M^2} {4 E^2} - \Phi}\ \cosh \omega - \frac M {2E},
\nonumber \\
s - s_B({\cal M}) &=& \frac {\ell} {\sqrt{2E}}\ \left(\sqrt{\frac {M^2} {4 E^2}
- \Phi}\ \sinh \omega - \frac M {2E}\ \omega\right),
\end{eqnarray}
where $\omega$ is the parameter, $\ell_2 = \pm 1$ and $s_B({\cal M})$ is an
arbitrary function of integration. In the L--T limit $Q = 0$, this function
becomes the bang-time. In the bang-free models, $s_B({\cal M})$ represents the
instant at which $R$ achieves a minimum or a maximum.

\begin{figure}
 \begin{center}
 \includegraphics[scale = 0.8]{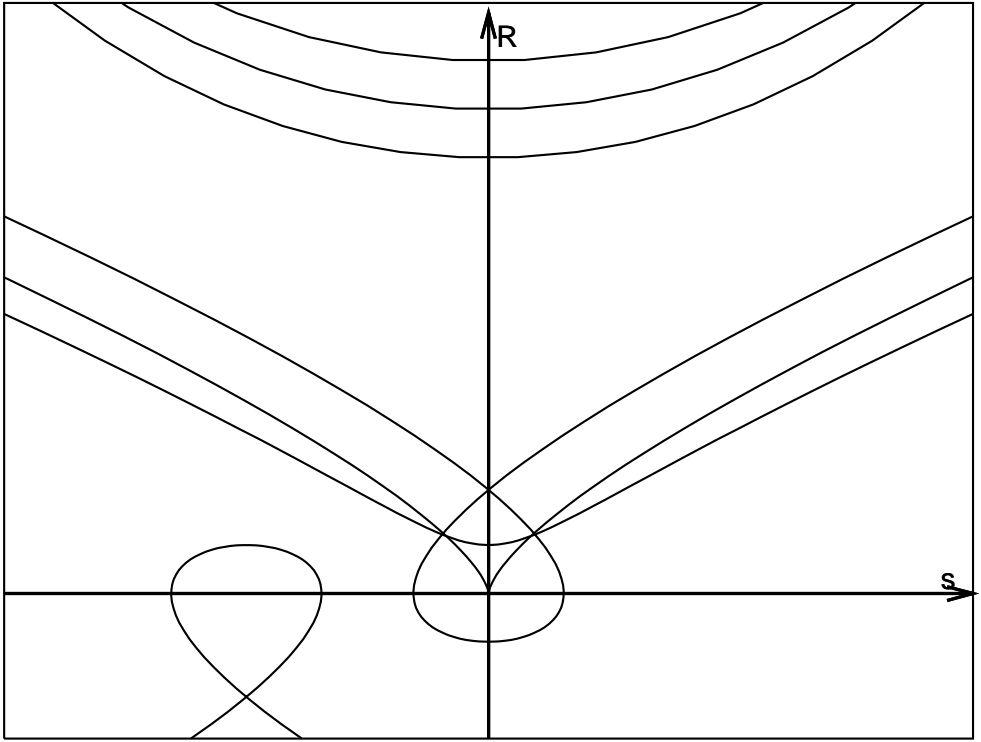}
 \caption{
 \label{chadu1fig}
 \footnotesize
Schematic graphs of the solutions (\ref{8.3}), with different values of $M$ and
$\Phi$. Only those parts of the curves that lie above the $s$ axis describe
physical situations. From bottom to top (counted at the left edge), the curves
correspond to: {\bf 1.} $\left\{\ell_2 = - 1, M < 0, \Phi > 0\right\}$. This
model has a finite time of existence. {\bf 2.} $\left\{\ell_2 = + 1, M > 0, \Phi
< 0\right\}$. {\bf 3.} $\left\{\ell_2 = + 1, M > 0, \Phi = 0\right\}$. This is
the uncharged (Lema\^{\i}tre -- Tolman) model, included for comparison. {\bf 4.}
$\left\{\ell_2 = + 1, M > 0, \Phi > 0\right\}$. The collapsing branch (left)
ends in a singularity at $R = 0$, the expanding branch (right) starts at the
singularity at $R = 0$. Models 1, 3 and 4 are the only ones with singularities.
{\bf 5.} $\left\{\ell_2 = + 1, M < 0, \Phi > 0\right\}$. {\bf 6.} $\left\{\ell_2
= + 1, M < 0, \Phi = 0\right\}$. {\bf 7.} $\left\{\ell_2 = + 1, M < 0, \Phi <
0\right\}$. The instant of time-symmetry is in general different on different
$M$-shells. It was made the same in the figure only for better readability.}
 \end{center}
 \end{figure}

The solution (\ref{8.3}) has no BB/BC singularity when $\ell_2 = +1$ and either
$M < 0$ or $\Phi < 0$. Schematic graphs of the solutions are shown in Fig.
\ref{chadu1fig}.

\begin{figure}
 \begin{center}
 \includegraphics[scale = 0.5]{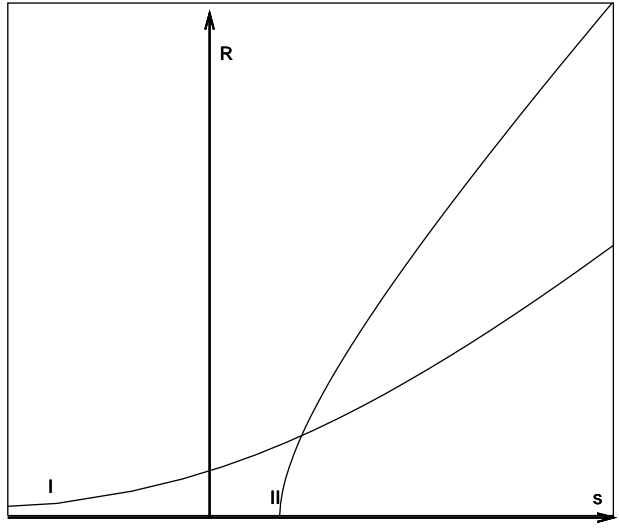}
 \caption{
 \label{chadu2fig}
 \footnotesize
Schematic graphs of the solutions (\ref{8.4}), with $\ell = +1$ and different
values of $M$. Curve I is the nonsingular solution with $M < 0$, curve II is a
solution with $M > 0$ and a singularity at finite $s$. In solution I $R$ tends
to zero asymptotically as $s \to -\infty$. }
 \end{center}
 \end{figure}

With $E > 0$ and $\Phi = M^2 / \left(4 E^2\right)$, the solution is
\begin{equation}\label{8.4}
s - s_B({\cal M}) = \frac {\ell} {2E} \left\{R - \frac M {2E}\ \ln
\left[\left.\left(R + \frac M {2E}\right)\right/ R_0\right]\right\}.
\end{equation}
This one will have no BB/BC singularity only if $M < 0$. Schematic graphs of
these solutions are shown in Fig. \ref{chadu2fig}.

\begin{figure}
 \begin{center}
 \includegraphics[scale = 0.5]{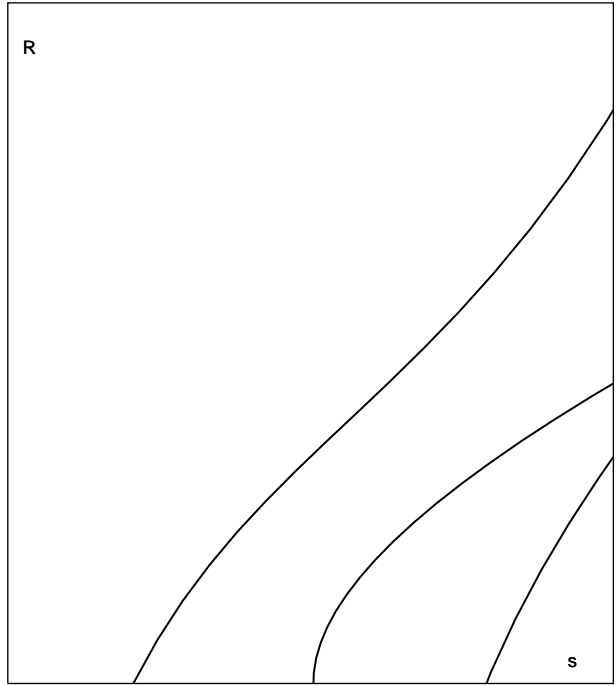}
 \caption{
 \label{chadu3fig}
 \footnotesize
Schematic graphs of the solutions (\ref{8.5}), with $\ell = +1$ and different
values of $M$ and $\Phi$. Every solution has a singularity; only the slope of
the curve at the singularity and the instant of the singularity depend on $M$
and $\Phi$. }
 \end{center}
 \end{figure}

With $E > 0$ and $\Phi > M^2 / \left(4 E^2\right)$, the solution is
\begin{eqnarray}\label{8.5}
R &=& \sqrt{\Phi - \frac {M^2} {4 E^2}}\ \sinh \omega - \frac M {2E}, \nonumber
\\
s - s_B({\cal M}) &=& \frac {\ell} {\sqrt{2E}}\ \left(\sqrt{\Phi - \frac {M^2}
{4 E^2}}\ \cosh \omega - \frac M {2E}\ \omega\right).
\end{eqnarray}
This model always runs into a BB/BC singularity at $\sinh \omega = M /
\sqrt{4E^2 \Phi - M^2}$, independently of the sign of $M$. It has no uncharged
limit $\Phi = 0$. Schematic graphs of the solutions are shown in Fig.
\ref{chadu3fig}.

With $E = 0$, we have to separately consider the subcase $M = 0$. Unlike in
neutral dust, this subcase is not vacuum (see (\ref{2.29})); it corresponds to
the case when the electrostatic repulsion among the dust particles exactly
balances the gravitational attraction, so that the effective mass is zero. This
subcase exists with all signs of $E$, but does not require separate treatment
when $E \neq 0$. (Such cases exist also in Newton's theory. The Newtonian
solutions with $M = 0$ and $E > 0$ are expanding or collapsing with constant
velocity, $R = \pm \sqrt{2E} \left(t - t_0\right) + R_0$, those with $M = 0 = E$
are static.)

\begin{figure}
 \begin{center}
 \includegraphics[scale = 0.8]{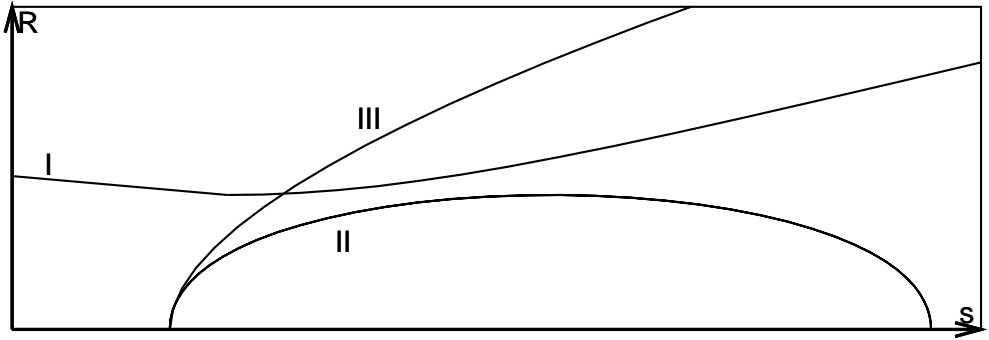}
 \caption{
 \label{chadu4fig}
 \footnotesize
Schematic graphs of the solutions (\ref{8.6}), with $\ell = \pm 1$ for curve I
and $\ell = +1$ for the other two curves, and with different values of $M$ and
$Q,_N$. Curve I is the nonsingular solution with $M > 0$ and ${Q,_N}^2 < G /
c^4$, curve II corresponds to $M < 0$ and ${Q,_N}^2 > G / c^4$, curve III
corresponds to $M > 0$ and ${Q,_N}^2 > G / c^4$. Model II achieves maximal
expansion at $R = - Q^2 \left({Q,_N}^2 - G / c^4\right) / (2M)$ and then
recollapses. }
 \end{center}
 \end{figure}

When $E = 0$ and $M \neq 0$, the solution is
\begin{equation}\label{8.6}
s - s_B({\cal M}) = \frac {\ell} {3M^2} \sqrt{2MR + Q^2 \left({Q,_N}^2 - G /
c^4\right)}\ \left[MR - Q^2 \left({Q,_N}^2 - G / c^4\right)\right],
\end{equation}
and it will not run into a BB/BC singularity only if $M > 0$ and ${Q,_N}^2 < G /
c^4$. Schematic graphs of the solutions are shown in Fig. \ref{chadu4fig}.

When $E = 0 = M$, the solution exists only when ${Q,_N}^2 > G / c^4$, and it is
\begin{equation}\label{8.7}
R = \sqrt{2 \ell |Q| \sqrt{{Q,_N}^2 - G / c^4}\ \left[s - s_B({\cal M})\right]}.
\end{equation}
This one always has a singularity. A plot has a similar shape as curve III in
Fig. \ref{chadu4fig}.

\begin{figure}
 \begin{center}
 \includegraphics[scale = 0.8]{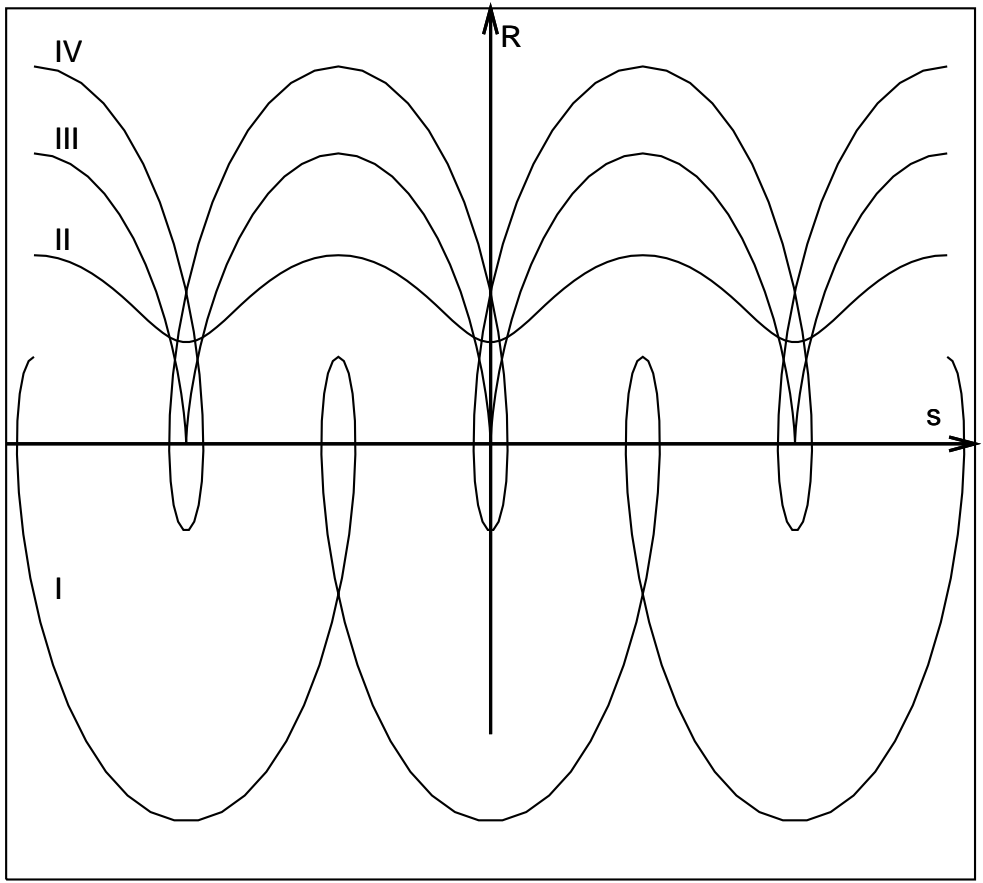}
 \caption{
 \label{chadu5fig}
 \footnotesize
Schematic graphs of the solutions (\ref{8.8}), with $\ell = +1$ and with
different values of $M$ and $\Phi$. Only those parts of the curves that lie
above the $s$-axis describe physical situations. The values of the parameters
are as follows: Curve I -- $M < 0$ and $\Phi < 0$; Curve II -- $M > 0$ and $\Phi
> 0$; Curve III -- $M > 0$ and $\Phi = 0$ (this is the Lema\^{\i}tre -- Tolman
evolution, shown for comparison); Curve IV -- $M > 0$ and $\Phi < 0$. Curve II
is the nonsingular model, all the other models begin and end with a singularity,
and the derivative $\dril R s$ is always infinite at the singularity. Different
$M$-shells have different periods in $s$. The periods were made equal only to
improve the readability of the picture. }
 \end{center}
 \end{figure}

When $E < 0$, a solution exists only with $\Phi < M^2 / \left(4E^2\right)$, and
it is
\begin{eqnarray}\label{8.8}
R &=& - \frac M {2E} - \sqrt{\frac {M^2} {4 E^2} - \Phi}\ \cos \omega, \nonumber
\\
s - s_B({\cal M}) &=& \frac {\ell} {\sqrt{- 2E}}\ \left(- \frac M {2E}\ \omega -
\sqrt{\frac {M^2} {4 E^2} - \Phi}\ \sin \omega\right).
\end{eqnarray}
This will avoid a BB/BC singularity only if $M > 0$ and $\Phi > 0$. Schematic
graphs of the solutions are shown in Fig. \ref{chadu5fig}. This is a periodic
solution, and the period is $T = 2 \pi M / (- 2E)^{3/2}$.

Note that if the functions $M(r)$, $E(r)$ and $Q(r)$ are chosen in accordance
with the regularity conditions of Sec. 6, then the regular behaviour of $R$ at
the center, $\lim_{r \to r_c} R/{\cal M}^{1/3} = \beta(t)$, automatically
follows in eqs. (\ref{8.3}) -- (\ref{8.6}) and (\ref{8.8}), but not in
(\ref{8.7}).

\section{Transition from collapse to expansion by nonsingular bounce}

\setcounter{equation}{0}

We will now investigate transferring the solution (\ref{7.1}) -- (\ref{7.2})
from the collapsing sector to the expanding sector. This is a nontrivial problem
because, as we have seen, the bounce point along each worldline, where $u^R =
0$, is a singularity of the mass-curvature coordinates, and we deduced
(\ref{7.1}) -- (\ref{7.2}) having $u^R < 0$ in mind. On the other hand, all
equations up to (\ref{5.18}) apply independently of the sign of $u^R$.

Let us rewrite eq. (\ref{2.23}) treating $({\cal M}, R)$ as the independent
variables and $t$ as the unknown function. Proceeding as in eqs. (\ref{5.6}) --
(\ref{5.7}), we find that (\ref{2.23}) is equivalent to
\begin{equation}\label{9.1}
\dr t R = \pm \frac {{\rm e}^{-C({\cal M}, R)/2}} {\sqrt{\Gamma^2 - 1 + 2M/R +
Q^2 \left({Q,_N}^2 - G / c^4\right) / R^2 - (1/3) \Lambda R^2}},
\end{equation}
where $+$ corresponds to expansion, $-$ to collapse, and $(\Gamma, M, Q)$ are
now functions of ${\cal M}$. The solution of (\ref{9.1}) is
\begin{equation}\label{9.2}
t = \pm J({\cal M}, R, R_0) + t_B({\cal M}),
\end{equation}
where $t_B({\cal M})$ is an arbitrary function and
\begin{equation}\label{9.3}
J({\cal M}, R, R_0) \df \int_{R_0}^R \frac {{\rm e}^{-C({\cal M}, w)/2} {\rm d}
w} {\sqrt{\Gamma^2 - 1 + 2M/w + Q^2 \left({Q,_N}^2 - G / c^4\right) / w^2 -
(1/3) \Lambda w^2}}
\end{equation}
The constant $R_0$ is an arbitrarily chosen initial point of integration, and
$t_B({\cal M})$ is the integration ``constant'', an analogue of the
Lema\^{\i}tre -- Tolman bang-time function. Here, $t_B({\cal M}) =
\left.t\right|_{R = R_0}$. Since $R$ is the only quantity that may depend on $t$
in the comoving coordinates, we see that it depends on $t$ always through the
combination $\left(t - t_B\right)$.

Note that eqs. (\ref{9.1}) -- (\ref{9.3}) apply only throughout a single
expansion phase or a single collapse phase. Thus, when we want to consider the
transfer from collapse to expansion in those models in which it is nonsingular,
we have to choose $R_0 = R_{\rm min}$, the minimal value achieved by $R$ for a
given ${\cal M}$, and thus $t = t_B$ will correspond to the bounce instant.

Consider now a pair of solutions (\ref{9.2}) -- (\ref{9.3}), the first one
collapsing, the other one expanding, both with the same function $t_B({\cal
M})$, and consider the two solutions at the same values of $({\cal M}, R, R_0)$.
In accordance with the previous paragraph, we have to choose $t_B({\cal M}) =
t({\cal M}, R_{\rm min})$ for both. Let the collapsing solution go through these
values of $({\cal M}, R, R_0)$ at $t = t_c$, and the expanding one at $t = t_e$.
We have $t_c = - J({\cal M}, R, R_0) + t_B$ for the collapsing solution, and
$t_e = + J({\cal M}, R, R_0) + t_B$ for the expanding one. Thus, $t_c + t_e =
2t_B$, i.e. $t_B({\cal M})$ is the midpoint of the interval $[t_c, t_e]$. In
general, it thus depends on ${\cal M}$, i.e. is different for each mass-shell.
The case $t_B =$ const, i.e. when the midpoint has the same value for all
masses, corresponds to the evolution that is time-symmetric with respect to $t =
t_B$.

Now consider eq. (\ref{5.10}) (recall: the $f$ in (\ref{5.10}) is the comoving
time $t$). Let us consider a pair of curves $(C_c, C_e)$ in the $(R, {\cal M})$
surface given by the same equation $R = R_{ce}({\cal M})$, one curve in the
collapsing region, the other one in the expanding region. Write eq. (\ref{5.10})
first for the collapsing solution, along $C_c$:
\begin{equation}\label{9.4}
\frac {{\rm e}^{C/2}} {\Gamma - QQ,_N/R}\ \left[- J,_{\cal M} ({\cal M}, R, R_0)
+ t_{B,{\cal M}} ({\cal M})\right] = I_c ({\cal M}, R, R_1) + g_c({\cal M}),
\end{equation}
where the functions $I$ and $g$ are those defined in (\ref{7.1}) -- (\ref{7.2}),
and the subscript $c$ refers to collapse. Write the same equation for the
expanding solution, along $C_e$:
\begin{equation}\label{9.5}
\frac {{\rm e}^{C/2}} {\Gamma - QQ,_N/R}\ \left[+ J,_{\cal M} ({\cal M}, R, R_0)
+ t_{B,{\cal M}} ({\cal M})\right] = I_e ({\cal M}, R, R_1) + g_e({\cal M}).
\end{equation}
The $I_e$ in (\ref{9.5}) differs from the $I_c$ in (\ref{9.4}) only by the sign
of $u^R$, otherwise all quantities in $I_e$ are the same as their counterparts
in $I_c$. Thus, along the chosen pair $(C_c, C_e)$, we have $I_c + I_e = 0$.
Knowing this, and adding eqs. (\ref{9.4}) and (\ref{9.5}), we obtain
\begin{equation}\label{9.6}
\frac {{\rm e}^{C/2}} {\Gamma - QQ,_N/R}\ 2t_{B, {\cal M}}({\cal M}) = g_c({\cal
M}) + g_e({\cal M}).
\end{equation}
This is the equation that relates $g({\cal M})$ for the collapsing solution in
(\ref{7.1}) -- (\ref{7.2}) to the corresponding function in the expanding
solution. For a time-symmetric evolution, and only then, we thus get $g_e = -
g_c$. Note that if the l.h.s. of (\ref{9.6}) is positive and large, while $g_c$
is positive and smaller than the l.h.s., then $g_e$ {\it cannot be negative}.
This observation will prove important below, in analysing the
existence/avoidance of shell crossings.

We recall (see the comment after (\ref{5.13})) that $F > 0$ where $\Gamma u^R <
0$ and $F < 0$ where $\Gamma u^R > 0$. As the bounce point is approached from
the $u^R > 0$ sector in the central region, $R$ eventually becomes smaller than
$R_1$, $F$ becomes unbounded, and it must tend to $- \infty$. We concluded
before that the expression in curly braces must be negative if shell crossings
are to be avoided in collapse. Thus, the first term in (\ref{7.1}) --
(\ref{7.2}) gives the correct behaviour of $F$ in approaching the bounce point
from the $u^R > 0$ sector.

The function $F$ is thus discontinuous at the bounce set, it tends to $+ \infty$
when this set is approached from the $u^R < 0$ sector, and to $- \infty$ when it
is approached from the $u^R > 0$ sector. But, as seen from eqs. (\ref{5.11}),
the metric components depend on $F^2$, $F / u^R$ and $\left(u^R\right)^2$, and
so they do not jump from one infinity to the other. Then, since $Fu^R$ is finite
at bounce and does not change sign, we see from (\ref{5.13}) that the mass
density is continuous across the bounce set.

Let us note that only the integral term $I$ in (\ref{7.1}) simply changes sign
when we go over from collapse to expansion. The function $g({\cal M})$
transforms by (\ref{9.6}), and it changes to its negative only for
time-symmetric evolution. Thus, if we have guaranteed that there are no shell
crossings before the bounce by choosing such $g$ that gives $I + g > 0$ in the
collapse phase, but the evolution is not time-symmetric ($t_{B, {\cal M}} \neq
0$), then $I + g < 0$ will in general fail to hold in the expansion phase, and
shell crossings will appear after the bounce. We shall come back to this
question later.

\section{Nonsingular bounce of a weakly charged dust}\label{nonbounce}

\setcounter{equation}{0}

From now on we assume ${Q,_N}^2 < G / c^4$ (weakly charged dust).

The cases $E \geq 0$ and $E < 0$ have to be considered separately. In the first
case, if a totally nonsingular model existed, it would be first collapsing from
an arbitrary initial density, then it would go through just one bounce, and
re-expand so that each non-central shell achieves an infinite radius. In the
second case, each shell would oscillate between the two turning values of $R$.
At each turning value, $u^R = 0$ and $|F| \to \infty$, and we have to choose the
arbitrary functions in (\ref{7.1}) -- (\ref{7.2}) so that the infinity has
always the right sign (e.g. $F \to + \infty$ when $u^R < 0$ and $\Gamma > 0$
before the turning point). Then, in principle we can secure the right sign in
two ways in each case: by imposing conditions on $\Gamma,_{\cal M}$ or on
\begin{equation}\label{10.1}
F_1 \df 1 - \left(c^4 / G\right) \left({Q,_N}^2 + QQ,_{NN}\right),
\end{equation}
depending on which one dominates in a given situation.

We first observe that a weakly charged nonsingular dust distribution with $E
\geq 0$ cannot exist. We assume that the regularity conditions at the center
hold, and that ${Q,_N}^2 \llim{{\cal M} \to 0} G / c^4$. Throughout the proof,
and then in deducing the conditions for no shell crossings in an $E < 0$ model,
we assume $u^R < 0$ (collapse), just for definiteness. In the end, it will be
easy to see that the same conditions must apply in the expansion phase.

Suppose, for the beginning, that $E = 0$ and consider collapse. Then $\Gamma =
\sqrt{2E + 1} = \pm 1$ is constant throughout the spacetime, and $\Gamma,_{\cal
M} \equiv 0$. Thus, $F_1$ must determine the sign of $F$ at the bounce point. To
secure the right sign also in the central region it must obey, in a vicinity of
the center, $F_1 < 0$, independently of the sign of $\Gamma$ (because $\Gamma F$
must have the same sign as long as $u^R$ does not change sign). In a vicinity of
the center the quantity $\left(1 - c^4 {Q,_N}^2 / G\right)$ is positive by
assumption, while ${Q,_N}^2 = G / c^4$ at the center. We have to consider two
cases:

1. Assume $Q,_N = + \sqrt{G} / c^2$ at the center. This means that $Q,_N > 0$ in
some vicinity of the center. Since, by the regularity conditions, $Q = 0$ at the
center, we have $Q > 0$ around the center. Then the condition $F_1 < 0$ means
$Q,_{NN} > \left( G / c^4 - {Q,_N}^2\right) / Q > 0$, which implies that $Q,_N$
in the vicinity of ${\cal M} = 0$ is larger than at ${\cal M} = 0$, i.e. larger
than $\sqrt{G} / c^2$. This is a contradiction with the assumption.

2. Assume $Q,_N = - \sqrt{G} / c^2$ at the center. Then $Q,_N < 0$ in some
vicinity of the center, and from the regularity conditions $Q < 0$ in the same
vicinity. Then $F_1 < 0$ implies $Q,_{NN} < \left( G / c^4 - {Q,_N}^2\right) / Q
< 0$, which means $Q,_N < - \sqrt{G} / c^2$ in a neighbourhood of the center, in
contradiction to the assumption.

Thus, $E = 0$ leads to a contradiction and is thereby excluded.

Suppose then that $E > 0$ and that $F_1 / (R\Gamma)$ becomes negligible as $x
\to 0$, so that the sign of $\Gamma,_{\cal M}$ determines the sign of infinity
of $F$ at bounce. We still consider collapse. Since, with $\Gamma > 0$ and $u^R
< 0$, $F$ must be positive, $\Gamma,_{\cal M} < 0$ is the right choice. The
regularity conditions require that $\Gamma^2 = 1$ at the center. If $\Gamma
> 0$, then $\Gamma = +1$ at the center, and then $\Gamma,_{\cal M} < 0$ means $0
< \Gamma < 1$ in the vicinity of the center. But $2E = \Gamma^2 - 1$, so this
implies $E < 0$. (With $\Gamma < 0$, the argument is similar: then $F < 0$, so
$\Gamma,_{\cal M} > 0$ and $\Gamma = -1$ at the center, which implies $E < 0$ in
the neighbourhood.)

Then suppose that $E > 0$ and that the limit at ${\cal M} \to 0$ of $F_1/R$ is
finite (which means that the limit of $F_1 / {\cal M}^{1/3}$ must be finite). As
$R \to \infty$, the contribution from $F_1/(x \Gamma)$ in (\ref{7.2}) becomes
negligible, and $R > R_1$ eventually. Thus, with $\Gamma
> 0$, $F$ must go to $+ \infty$ as $R \to \infty$, so $\Gamma,_{\cal M}
> 0$. In the vicinity of the center, where $R < R_1$, we must have
$\Gamma,_{\cal M} + F_1/(R\Gamma) < 0$, so $F_1 < 0$. The condition $F_1 < 0$
also follows when $\Gamma < 0$. Thus $F_1 / {\cal M}^{1/3} < 0$ at the center
and, by continuity, in some neighbourhood of the center. Then we use the same
argument as above to show that it leads to a contradiction in every case.

In the case $E < 0$, these problems can be avoided. With $E < 0$, there are two
turning points, at $R = R_+$ and at $R = R_- > R_+$. The integral in (\ref{7.2})
becomes unbounded at both $R_+$ and $R_-$. The signs of the two infinities can
then be imposed independently. The argument used after (\ref{10.1}) still
applies, so $F_1$ cannot secure the correct sign at $R_+$. Thus we have to
assume that the limit of $F_1/R$ as ${\cal M} \to 0$ is zero, so that
$\Gamma,_{\cal M} < 0$ secures the right behaviour at the inner turning point.
Then we require that at $R_-$ the inequality $1 - \left(c^4 / G\right)
\left({Q,_N}^2 + QQ,_{NN}\right) > - \Gamma \Gamma,_{\cal M} R_-$ holds, which
reads explicitly:
\begin{equation}\label{10.2}
1 - \frac {c^4} G \left({Q,_N}^2 + QQ,_{NN}\right) > \frac {\Gamma \Gamma,_{\cal
M}} {2E}\ \left[M + \sqrt{M^2 - 2EQ^2 \left({Q,_N}^2 - G / c^4\right)}\right].
\end{equation}
Since $\Gamma,_{\cal M} < 0$ and $E < 0$ in a neighbourhood of ${\cal M} = 0$,
this implies the following necessary condition;
\begin{equation}\label{10.3}
1 - \frac {c^4} G \left({Q,_N}^2 + QQ,_{NN}\right) > \frac {\Gamma \Gamma,_{\cal
M}} {2E}\ M.
\end{equation}
Since the integral in (\ref{7.2}) goes to $+\infty$ at both turning points, it
must have at least one local minimum somewhere in $\left(R_+, R_-\right)$. If $-
\infty < I_{\rm min} < 0$ at the smallest of these minima, then this can be
corrected by the choice of $g({\cal M})$. Also, $I$ equals to zero in at least
one point -- at $R = R_1$, where $R_+ < R_1 < R_-$. Thus, $I$ cannot be positive
everywhere, so necessarily $g({\cal M}) > 0$ to avoid $F = 0$ anywhere.

In summary, a weakly charged configuration that could bounce singularity-free
through the R--N wormhole must obey the following necessary conditions in a
neighbourhood of the center:

(1) $E < 0$ (as $E \geq 0$ was eliminated above);

(2) $E \geq -1/2$ (since $\Gamma$ is the primary arbitrary function, and $E$ was
defined as an auxiliary quantity by $2E = \Gamma^2 - 1$)\footnote{The condition
$E \geq -1/2$ finds a clearer explanation in the Lema\^{\i}tre -- Tolman (zero
charge) limit, where it follows from the requirement of the right signature. It
is also equivalent to the statement that there can be no turning points for
radial motion of uncharged dust inside the apparent horizon.};

(3) $\lim_{r \to r_c} F_1/{\cal M}^{1/3} = 0$, in consequence of $\lim_{r \to
r_c} F_1/R = 0$ and of (\ref{6.2}) (recall that $\beta \neq 0$ in (\ref{6.2})).

(4) $\Gamma,_{\cal M} < 0$ (to secure the right sign of $F$ at the inner turning
point);

(5) ${Q,_N}^2 < G / c^4$ at $N > 0$ and ${Q,_N}^2 = G / c^4$ at $N = 0$ (the
defining condition for weakly charged dust);

(6) $M \equiv {\cal M} - QQ,_N \Gamma > 0$ (a necessary condition for no BB/BC
with $E < 0$, as shown in Secs. 4 and 8);

(7) $M^2 - 2E Q^2 \left({Q,_N}^2 - G / c^4\right) > 0$ (a necessary condition
for the existence of a solution of (\ref{2.23}), see (\ref{4.2}); equality leads
to a static solution);

(8) condition (\ref{10.3}) (a necessary condition for (\ref{10.2}) to be
obeyed);

(9) condition (\ref{10.2}) (to secure the right sign of $F$ at the outer turning
point).

\noindent In addition to that, all the regularity conditions of sec. 6 must be
obeyed.

Conditions (1) -- (9) must hold in a {\it neighbourhood} of the center. {\it At}
the center, the left-hand sides of conditions (1) and (6 -- 9) must have zero
limits.

Condition (\ref{10.2}) should hold at the state of maximal expansion. However,
for those shells that contain small mass (are sufficiently close to the center),
the value of $R_-$ will be arbitrarily small. Thus (\ref{10.2}) must in fact
hold down to arbitrarily small values of ${\cal M}$, with only ${\cal M} = 0$
being excluded, where both sides of (\ref{10.2}) should be zero.

Now, a glance at eqs. (\ref{7.1}) -- (\ref{7.2}) suffices to see that the same
conditions (1) -- (9) must apply in the expansion phase, where $u^R > 0$ and
$\Gamma F < 0$, if shell crossings are to be avoided. The two functions $g$ are
related to each other by (\ref{9.6}), and it may happen that if we guaranteed $F
= I_c + g_c > 0$ in collapse, then $I_e + g_e = - I_c + g_e$ will refuse to be
negative in some range during expansion, thus indicating a shell crossing. But
if the bounce is time-symmetric, so that $g_e = - g_c$, then, having guaranteed
no shell crossings in collapse, we know that there will be no shell crossings in
the subsequent expansion phase. However, the next bounce will in general no
longer be time-symmetric, and shell crossings will appear in later cycles. The
shell crossings might be avoided for ever only if {\it all bounces} were
time-symmetric. Equation (\ref{8.8}) implies that such models might exist: the
period of oscillations $T = 2\pi M/(-2E)^{3/2}$ will be independent of ${\cal
M}$ when $M/(-2E)^{3/2}$ is constant. Together with $t_B$ being constant, such a
condition reduces the Lema\^{\i}tre -- Tolman model to the Friedmann limit
\cite{Kras1997}, but here we still have the function $Q$ that generates the
electric field. With $T = T_0 =$ const, the function $Q(N)$, in general
arbitrary, has to obey
\begin{equation}\label{10.4}
QQ,_N = \frac G {c^4 {\cal M},_N}\ \left[{\cal M} - \frac {T_0} {2\pi} \left(1 -
\Gamma^2\right)^{3/2}\right],
\end{equation}
and it can still obey the regularity conditions $Q = 0$ and $\left|Q,_N\right| =
\sqrt{G}/c^2$ at the center, if ${\cal M}$ and $\Gamma$ obey their regularity
conditions. Also, (\ref{10.4}) does not lead to any simple contradiction with
conditions (5 -- 9). An explicit example of a configuration obeying (\ref{10.4})
(or a proof that it does not exist) remains to be found.

Equation (\ref{10.4}) is a consequence of the requirement that the period with
respect to the proper time $s$ is independent of ${\cal M}$. It is easy to find
from (\ref{9.1}) that the period with respect to the time coordinate $t$ (equal
to twice the integral of (\ref{9.1}) from $R_{\rm min} = R_+$ to $R_{\rm max} =
R_-$) is $2\pi {\rm e}^{- C({\cal M}, \overline{R}({\cal M}))/2} M/(-
2E)^{3/2}$, where $\overline{R}({\cal M})$ is an intermediate value of $R$
between $R_{\rm min}$ and $R_{\rm max}$ (from the mean value theorem for
integrals). Thus, the constancy of the period in $t$ also imposes an additional
equation on $Q$. Here, however, an example could be found only numerically.

It should be noted that $E$ need not have the same sign in the whole volume of
the charged dust. However, the change of sign of $E$ will have no influence on
the conditions stated above. If it gets positive in a vicinity of the center,
then nonsingular bounce cannot occur, as shown above. If it gets negative in a
vicinity of the center, then it is negative in an open range of values of ${\cal
M}$. Then, within the same open set there will be outer turning points of some
mass-shells, so (\ref{10.3}) and (\ref{10.2}) must anyway hold close to the
center. However, if $E$ becomes positive at some ${\cal M} = {\cal M}_0$, then
there are no outer turning points at ${\cal M} \geq {\cal M}_0$, and the
inequalities (\ref{10.2}) and (\ref{10.3}) need not be imposed in that region.

\begin{figure}
 \begin{center}
 \includegraphics[scale = 0.8]{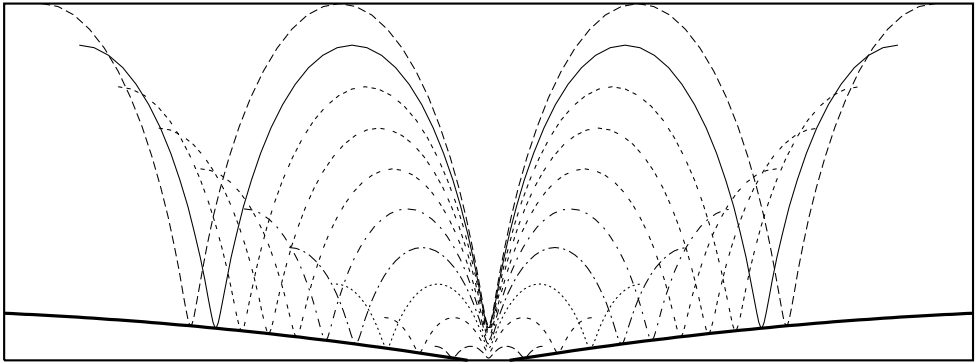}
 \caption{
 \label{cyclesfig}
 \footnotesize
A possible pathology of an oscillating model with $E < 0$: the period of
oscillations tends to zero as the center of symmetry is approached (the curves
are graphs of the functions $R(t, {\cal M})$ for different values of ${\cal
M}$). The thicker curve goes through the minima of the evolution graphs. (The
model shown in the figure is time-symmetric with respect to the bounce in the
middle of the axis.) }
 \end{center}
 \end{figure}

We have not yet excluded one more pathological situation that could occur with
$E < 0$. The arbitrary functions may be chosen so that the period of
oscillations tends to zero as ${\cal M} \to 0$, as in Fig. \ref{cyclesfig}. The
thicker curve (call it $E_{\cal M}$) goes through the minima of the evolution
curves $R(t, {\cal M})$, each of them corresponding to a fixed value of ${\cal
M}$. In the figure, the slope $\alpha({\cal M})$ of $E_{\cal M}$ is at each
point smaller than the slope of the evolution curve that passes through the same
point, and the shell crossings are inevitable. We could try to avoid it by
requiring that $\alpha({\cal M})$ is sufficiently large, but this would not
work: the period of a curve with larger ${\cal M}$ would be larger than the
period of a curve with smaller ${\cal M}$, and an intersection would be sure to
occur after several cycles. With the period tending to zero as ${\cal M} \to 0$,
the neighbourhood of the center ${\cal M} = 0$ would be densely filled with
shell crossings. Thus, the only way to avoid such a pathology is to choose the
arbitrary functions so that the period tends to a nonzero value as ${\cal M} \to
0$.

Equation (\ref{8.8}) will help us in choosing the right functions. We assume
that nonzero intervals of proper time along each worldline are mapped into
nonzero intervals of the comoving time-coordinate and vice versa (i.e. that the
metric function ${\rm e}^C$ in does not tend to zero or to infinity at the axis
${\cal M} = 0$). Thus, if we choose the functions so that the period in $s$ in
eq. (\ref{8.8}) is nonzero, then the period in $t$ will also be nonzero. Note
then that if $E$ obeys (\ref{6.4}) with a nonzero constant, then, in consequence
of condition (5), the period of $R$ in $s$ will tend to zero as ${\cal M} \to
0$. To avoid this, $E$ must tend to zero at the center faster than ${\cal
M}^{2/3}$.

\section{An example}\label{Anexample}

\setcounter{equation}{0}

In order to prove that conditions (1) -- (9) from the previous section are not
mutually contradictory, we shall now provide an example of a model that obeys
them. It is simply the first model that has successfully passed all the tests,
and we do not claim that it is physically important or realistic -- it is only
meant to be a proof of existence.

Since the functions appearing in the inequalities are rather complicated, the
proof that the inequalities are all obeyed will be given mainly by numerical
graphs. The connection between $\Gamma$, $N$ and ${\cal M}$ provided by
(\ref{5.16}) causes that even quite simple functions $E(N)$ that obey the
regularity conditions lead to complicated expressions for ${\cal M}(N)$.
However, we will verify by exact methods that in a vicinity of the center and at
infinity the functions indeed behave in the same way as the graphs indicate.

Choose, in agreement with the regularity conditions and the weak-charge
conditions:
\begin{equation}\label{11.1}
Q(N) = q \frac {\sqrt{G} N_0} {c^2}\ p(x),
\end{equation}
where $q = \pm 1$, to allow for any sign of the charge, $x \df N/N_0$, $N_0$ is
a constant, and
\begin{equation}\label{11.2}
p(x) =  x / (1 + x)^2.
\end{equation}
Then
\begin{equation}\label{11.3}
Q,_N = q \frac {\sqrt{G}} {c^2}\ \frac {1 - x} {(1 + x)^3}.
\end{equation}
This charge is zero at $x = 0$, its absolute value increases at first, but then
decreases and tends to zero as $x \to \infty$. Thus $Q,_N$, which is
proportional to the charge density (see (\ref{2.20})), changes sign at $x = 1$,
but it obeys condition (5) everywhere.

\begin{figure}
 \begin{center}
 \includegraphics[scale = 0.9]{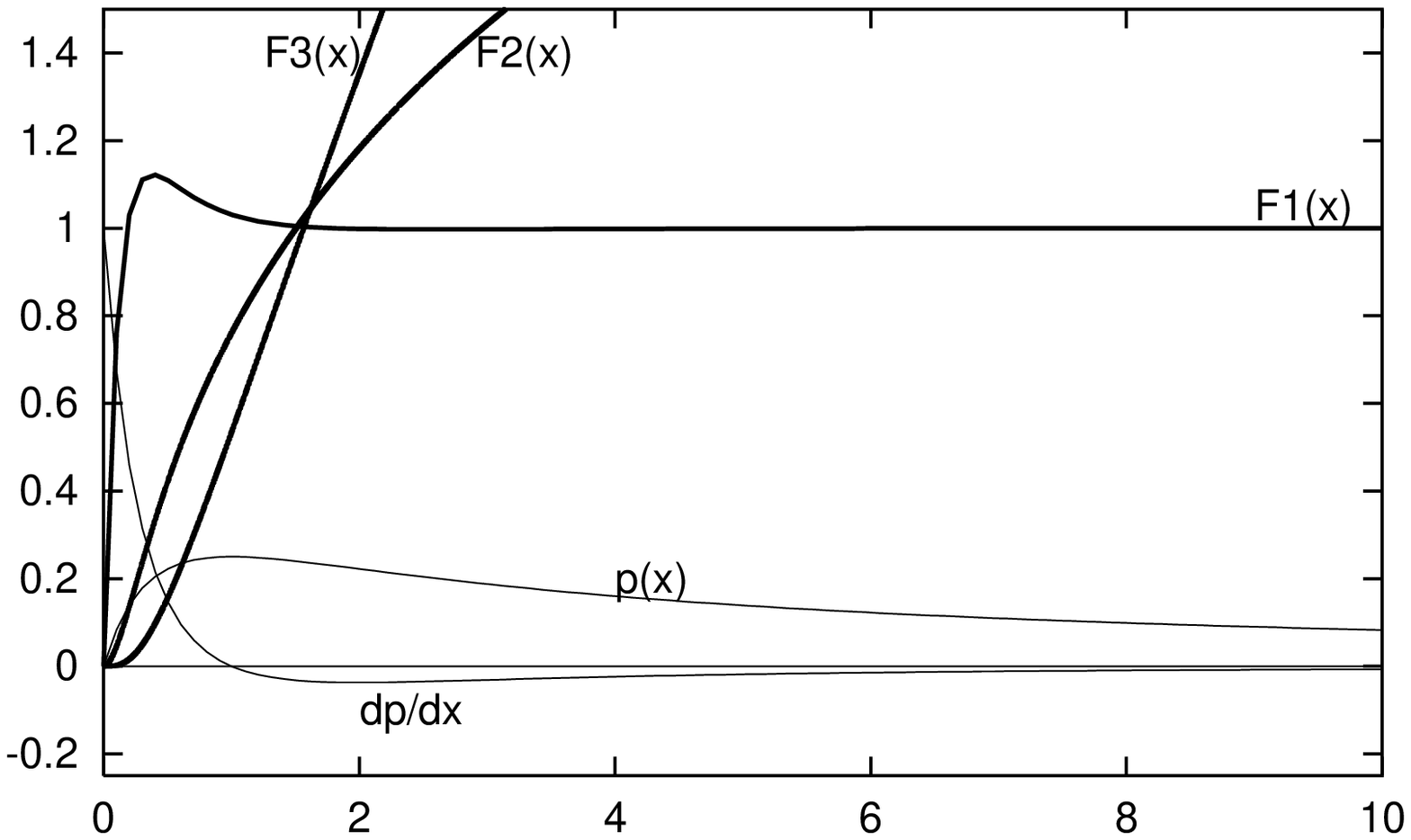}
 ${}$ \\[-45mm]
 \hspace{70mm}
  \includegraphics[scale = 0.35]{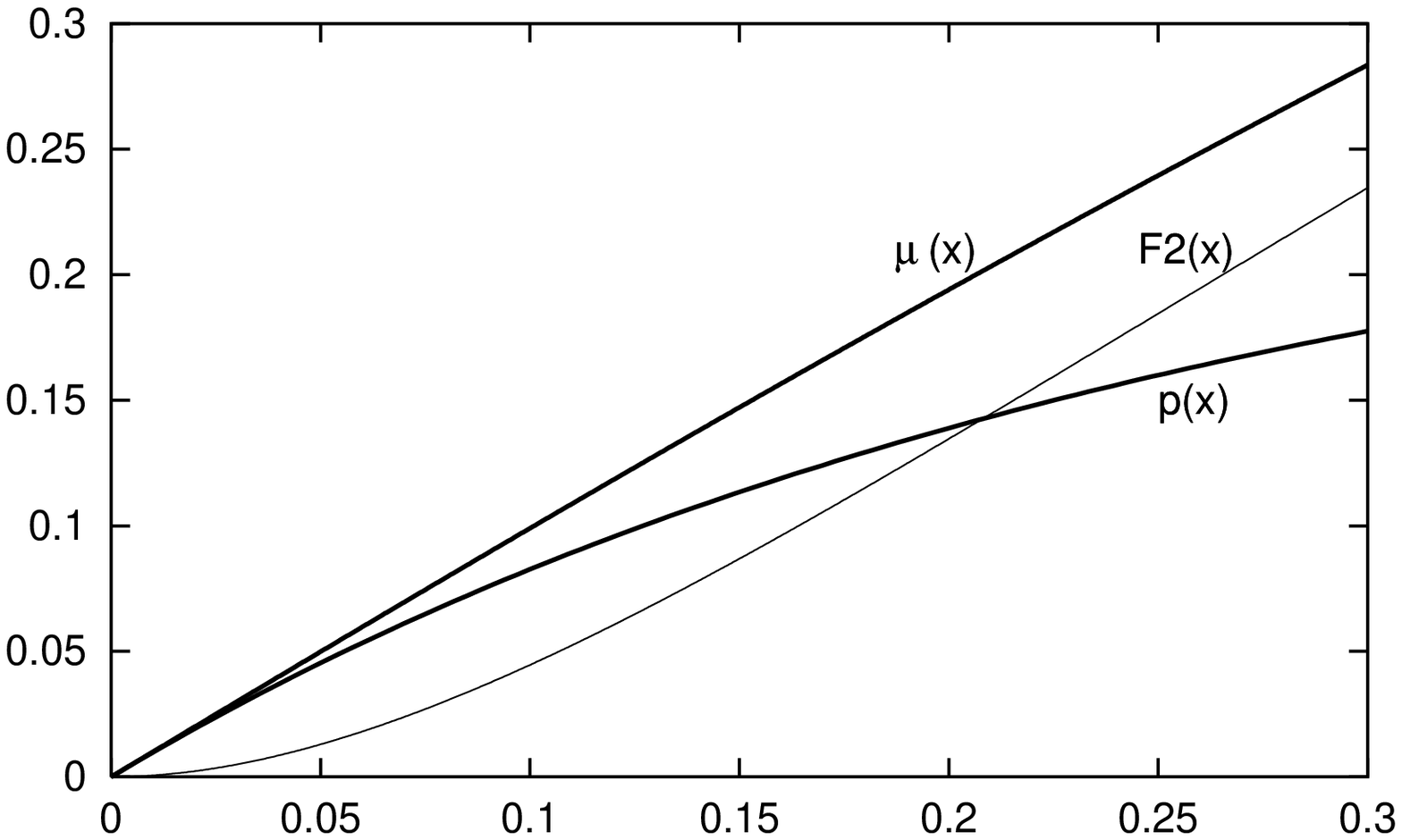}
 ${}$ \\[20mm]
 \caption{
 \label{chadu6fig}
 \footnotesize
Graphs of the functions $p(x)$, $\dril p x$, $F_1(x)$, $F_2(x)$ and $F_3(x)$.
Note that $-1 < \dril p x < +1$ and $F_2(x) > 0$, $F_3(x) > 0$ for all $x > 0$,
thus conditions (5), (6) and (7) are fulfilled. All the functions are zero at $x
= 0$. As $x \to \infty$, $F_2$ and $F_3$ tend to $\infty$, $F_1$ tends
asymptotically to 1, while $p$ and $\dril p x$ tend asymptotically to 0. The
inset shows the functions $p(x)$, $F_2(x)$ and $\mu(x)$ in a vicinity of $x =
0$. It shows that, with our chosen value $b = 2.5$, $\mu(x) > p(x)$ for $x >
0$.}
 \end{center}
 \end{figure}

Further, we have
\begin{equation}\label{11.4}
F_1(x) \df 1 - \frac {c^4} G\ \left({Q,_N}^2 + QQ,_{NN}\right) = 1 - \frac {3x^2
- 6x + 1} {(1 + x)^6}.
\end{equation}
Fig. \ref{chadu6fig} shows the shape of the functions $p(x)$, $\dril p x$ and
$F_1(x)$ (for all these functions it can be easily verified that the graphs show
their behaviour faithfully, with no important details being hidden beyond the
range of the graph or in a small vicinity of zero).

We choose now the function $E(N)$ so that the period $2\pi M / (-2E)^{3/2}$ in
(\ref{8.8}) has a nonzero limit at $N \to 0$. In order to keep $-1/2 \leq E < 0$
in the whole range, and to avoid $\Gamma$ and ${\cal M} = (G/c^4) \int
(1/\Gamma) {\rm d} N$ being too complicated, we choose the trial form $2E = -
bx^a / (1 + bx^a)$, with $a$, $b$ being constant, and we find from (\ref{8.8})
that the period will have a nonzero (actually, infinite) limit at $N = 0$ if $a
= 5/3$. This will also guarantee, via (\ref{8.8}), that the limit of $R/{\cal
M}^{1/3}$ at $x \to 0$ will be a nonzero function independent of ${\cal M}$,
just as (\ref{6.2}) requires. Thus
\begin{equation}\label{11.5}
2E = - \frac {b x^{5/3}} {1 + b x^{5/3}}.
\end{equation}
With such $E$, the limiting period of oscillations in (\ref{8.8}) is infinite,
and
\begin{equation}\label{11.6}
\Gamma(x) = \frac 1 {\sqrt{1 + b x^{5/3}}},
\end{equation}
which obeys condition (4), and, further:
\begin{equation}\label{11.7}
{\cal M}(x) = \frac {GN_0} {c^4} \int_0^x \frac {{\rm d} x'} {\sqrt{1 + b
{x'}^{5/3}}} \df \frac {GN_0} {c^4} \mu(x).
\end{equation}
The $\mu(x)$ is an increasing function for all $x > 0$; its graph would almost
coincide with the graph of $F_2$ shown in Fig. \ref{chadu6fig}, but see the
inset (actually, $\mu(x) > F_2(x)$ for $0 < x < 1$, the two curves intersect at
$x = 1$, and for $x > 1$ $\mu(x) < F_2(x)$). Note that such ${\cal M}$ obeys
$\lim_{x \to 0} {\cal M}/x = GN_0/c^4 \neq 0$, while $\lim_{x \to 0} F_1/x =
12$, so condition (3) is obeyed. We have now:
\begin{eqnarray}\label{11.8}
M &\equiv& {\cal M} - QQ,_N \Gamma = \frac {GN_0} {c^4}\ F_2(x), \nonumber \\
F_2(x) &\df& \mu(x) - \frac {x (1 - x)} {(1 + x)^5 \sqrt{1 + b x^{5/3}}}.
\end{eqnarray}
It is easy to verify that $F_2 > 0$ for all $x > 0$,\footnote{We have $F_2(x)
> \mu(x) - x / \left[(1 + x)^5 \sqrt{1 + b x^{5/3}}\right] \df \widetilde{F}_2$
for all $x > 0$, and $\dril {\widetilde{F}_2} x > 0$ for all $x > 0$, so $F_2
> {\widetilde{F}_2} > 0$ for all $x > 0$.} so condition (6) is fulfilled. The
graph of $F_2$ is shown in Fig. \ref{chadu6fig} with $b = 2.5$ (why this value
-- see below).

Condition (7) is equivalent to
\begin{equation}\label{11.9}
F_3(x) > 0, \qquad F_3(x) \df {F_2}^2(x) - F_6(x),
\end{equation}
where
\begin{equation}\label{11.10}
F_6(x) \df - 2E p^2 \left(1 - {p,_x}^2\right) = \frac {b x^{11/3}} {\left(1 + b
x^{5/3}\right) (1 + x)^4}\ \left[1 - \frac {(1 - x)^2} {(1 + x)^6}\right].
\end{equation}
The graph of $F_3(x)$ with $b = 2.5$ is shown in Fig. \ref{chadu6fig}. Since we
have verified that $M \geq 0$, $E < 0$ and ${Q,_N}^2 \leq G/c^4$, equation
(\ref{11.9})) is equivalent to $\widetilde{F}_3 > 0$, where $\widetilde{F}_3(x)
\df F_2(x) - \sqrt{F_6(x)}$. Note that $\widetilde{F}_3(0) = 0$. We divide
$\widetilde{F}_3(x)$ by $x^2$ and observe that the second term in the resulting
expression is zero at $x = 0$, while $\lim_{x \to 0} F_2(x)/x^2 = 6 > 0$. Thus,
$\widetilde{F}_3 > 0$ in a vicinity of $x = 0$, and, similarly to $F_2(x)$,
tends to $+ \infty$ as $x \to \infty$.

\begin{figure}
 \begin{center}
 \includegraphics[scale = 0.8]{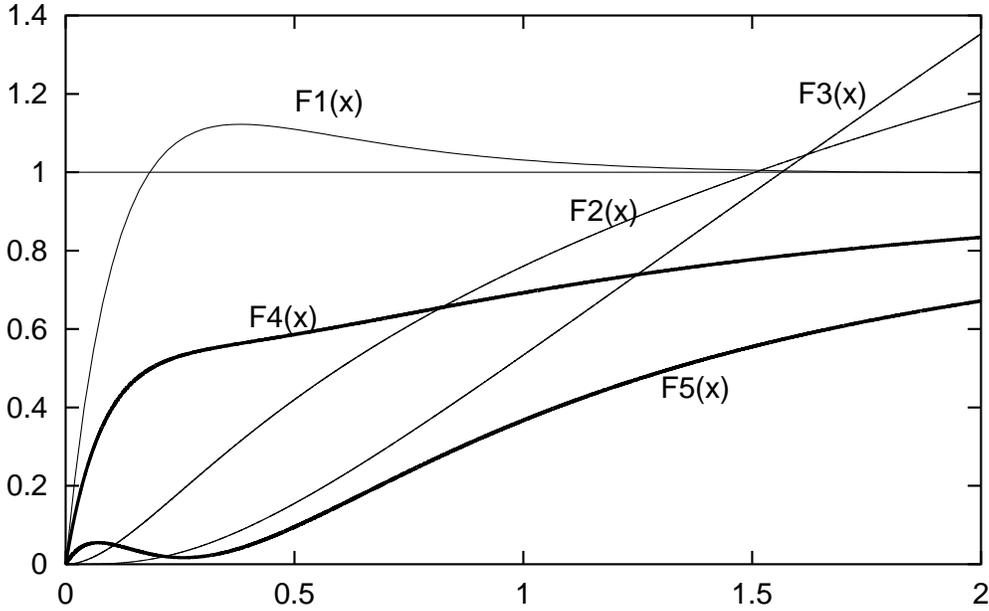}
 \caption{
 \label{chadu7fig}
 \footnotesize
Graphs of the functions $F_4(x)$ and $F_5(x)$ with $b = 2.5$. The graphs of
$F_1(x)$, $F_2(x)$ and $F_3(x)$ are shown for comparison and for scale. Both
$F_4(x)$ and $F_5(x)$ asymptotically tend to 1 as $x \to \infty$ and both have
vertical tangents at $x = 0$.}
 \end{center}
 \end{figure}

Since $\Gamma,_{\cal M} = \Gamma,_N N,_{\cal M}$, after using (\ref{5.16}), we
obtain:
\begin{equation}\label{11.11}
\frac {\Gamma \Gamma,_{\cal M}} {2E} = \frac {c^4} {GN_0}\ \frac 5 {6x \sqrt{1 +
b x^{5/3}}},
\end{equation}
and so condition (8) becomes:
\begin{equation}\label{11.12}
F_4(x) > 0, \qquad {\rm where} \qquad F_4(x) \df F_1(x) - \frac 5 {6x \sqrt{1 +
b x^{5/3}}}\ F_2(x).
\end{equation}
We have $F_4(0) = 0$. Knowing that $\lim_{x \to 0} F_2/x^2 = 6$, we easily
calculate that $\left.\dril {F_4} x\right|_{x = 0} = 7 > 0$. Thus $F_4(x)$ is
increasing in a vicinity of $x = 0$, i.e. $F_4(x) > 0$ at least for some $x >
0$. Also, $F_4(x) \llim{x \to \infty} 1$. The graph of $F_4(x)$ with $b = 2.5$
is shown in Fig. \ref{chadu7fig}.

Finally, condition (9), i.e. eq. (\ref{10.2}), can be written as
\begin{equation}\label{11.13}
F_5(x) > 0, \qquad {\rm where}\ F_5(x) \df F_1(x) - \frac 5 {6 x \sqrt{1 + b
x^{5/3}}} \left[F_2(x) + \sqrt{F_3(x)}\right],
\end{equation}
(see eq. (\ref{11.10}) for the definition of $F_6(x)$). Since we have already
verified that (\ref{11.12}) holds, we can take the term with the square root in
(\ref{11.13}) to the r.h.s. in $F_5 > 0$ and then square both sides of the
inequality. The result can be rewritten as:
\begin{equation}\label{11.14}
F_1 \left(F_1 - \frac 5 {3 x \sqrt{1 + b x^{5/3}}}\ F_2\right) + \frac {25} {36
x^2 \left(1 + b x^{5/3}\right)}\ F_6 > 0.
\end{equation}
We note that $\lim_{x \to 0} F_1/x = 12$, $\lim_{x \to 0} F_2/x^2 = 6$ and
$\lim_{x \to 0} F_6/x^{14/3} = 8b$. Then we divide (\ref{11.14}) by $x^2$ and
take the limit of the resulting expression at $x \to 0$. The result is $24 > 0$.
Hence, (\ref{11.14}) is obeyed in a vicinity of $x = 0$, and, consequently, $F_5
> 0$ is obeyed in that vicinity, too. It is easy to verify that $F_5(x) \llim{x
\to \infty} 1$. The graph of $F_5$ with $b = 2.5$ is shown in Fig.
\ref{chadu7fig}.

The reason for choosing $b = 2.5$ was this: the graph of $F_5$ is sensitive to
the value of $b$. With decreasing $b$, the local minimum of $F_5$ becomes
smaller, and for $b$ small enough (for example, $b = 0.75$) $F_5 < 0$ around the
minimum. With $b \geq 2.5$, the minimum is clearly positive.

Graphs do not constitute a definitive proof that a function is positive in the
whole infinite range. Fine details can be hidden beyond the range of the figures
or at a very small scale around the points where the functions approach zero. In
those regions the functions can go below zero, while we cannot see it. However,
we have verified by exact methods that there exists a neighbourhood $U$ of $x =
0$ in which the functions $F_1, \dots, F_5$ are positive for $x > 0$. Thus, even
if some of the functions become negative outside $U$, we can cut away a finite
ball of the charged dust with the radius $x_0$ smaller than the radius of $U$
and match the charged dust to the Reissner -- Nordstr\"{o}m solution at $x =
x_0$. In this way, we will obtain a finite charged body of dust that can go
through a minimal size and bounce without encountering any singularity.

\begin{figure}
 \begin{center}
 \includegraphics[scale = 0.8]{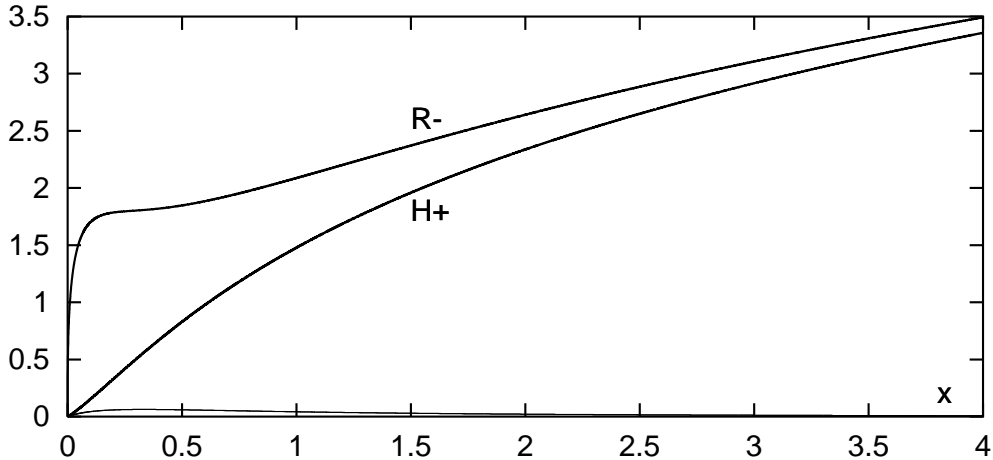}
 \caption{
 \label{plothors}
 \footnotesize
The maximal radius achieved in each cycle, $R_-$, and the radius of the outer
R--N horizon, $R_{H+}$, as functions of mass. The numbers on the axes are
multiples of $GN_0/c^4$. At this scale, the radius of the inner horizon and the
minimal radius seem to coincide with the $x$-axis, but see Fig.
\ref{plothormin}. As required by the equations of motion, we have $R_-
> R_{H+} > R_{H-} > R_+$ for all masses. Note that $R_-$ and $R_{H+}$ are
increasing all the way.}
 \end{center}
 \end{figure}

\begin{figure}
 \begin{center}
  \includegraphics[scale=0.8]{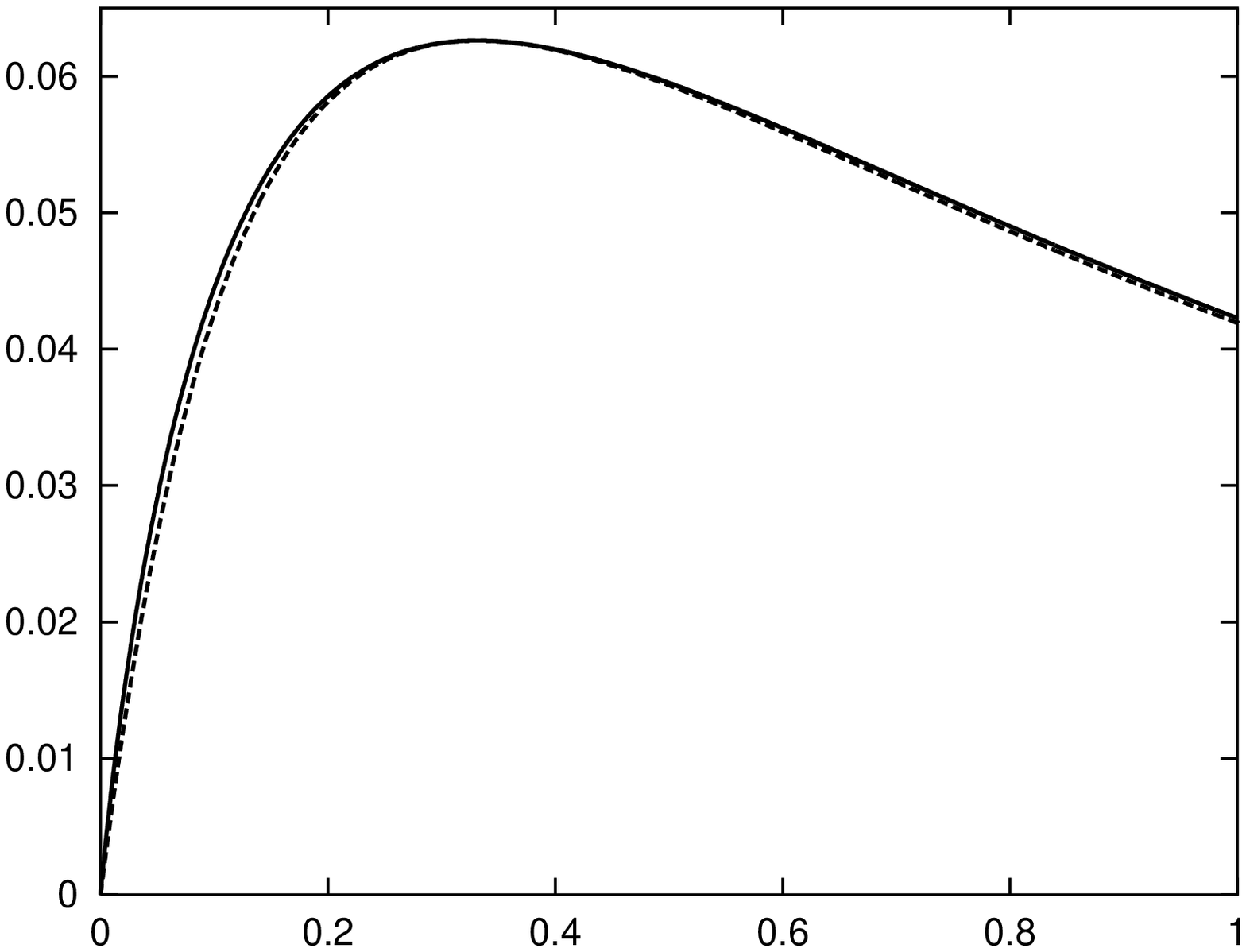}
  ${}$ \\[-60mm]
  \includegraphics[scale = 0.35]{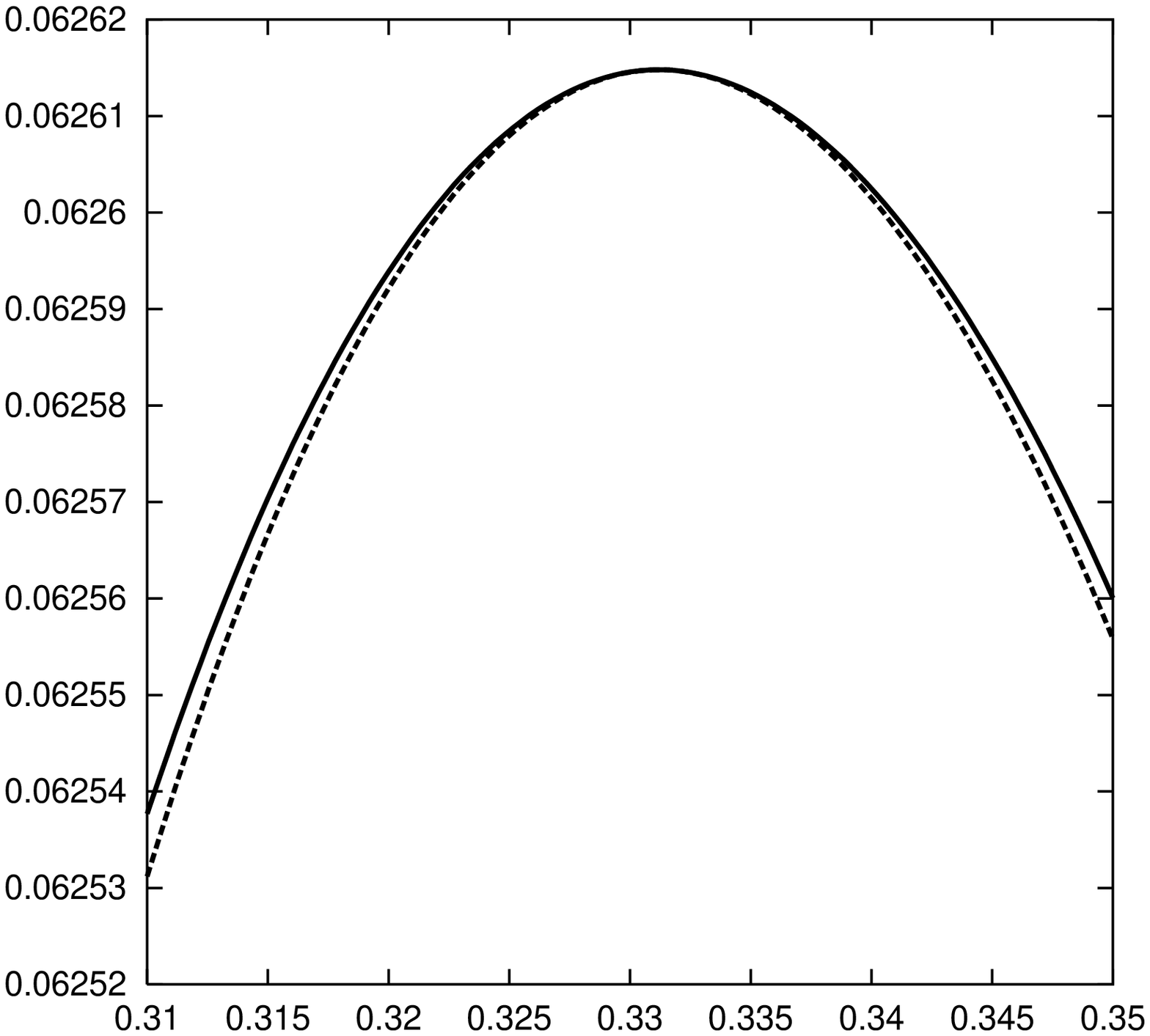}
  ${}$ \\[15mm]
 \caption{
 \label{plothormin}
 \footnotesize
The radius of the inner horizon, $R_{H-}$ (upper curve), and the minimal radius,
$R_+$ (lower curve), as functions of mass. The numbers on the axes are multiples
of $GN_0/c^4$. Both radii first increase with mass, but then decrease. The
figure shows that $R_+ < R_{H-}$ everywhere except at the maximum, where the two
curves are tangent to each other. The inset is a closeup view of the
neighbourhood of the maximum.}
 \end{center}
 \end{figure}

It can be easily verified that in our example $(G / c^4) Q^2 < {\cal M}^2$ for
all $x > 0$ provided $b < 25.3$ (hint: $p,_x < \mu,_x$ is equivalent to
$\left[(1 + x)^6/(1 - x)^2 - 1\right]^{3/2} > b^{3/2}x^{5/2}$ -- see the inset
in Fig. \ref{chadu6fig}). Thus, if a finite sphere is cut out of this
configuration and matched to the Reissner -- Nordstr\"{o}m solution, the
exterior R--N metric will have $e^2 < m^2$, and horizons will exist in
it.\footnote{But $GQ^2/(c^4 {\cal M}^2) \to 1$ as $x \to 0$, so as the sphere
becomes smaller, the outer R--N metric tends to the critical one, $e^2 = m^2$.}
As is well-known from the studies of the Reissner--Nordstr\"{o}m metric
\cite{PlKr2006}, the surface of a charged dust sphere matched to the R--N metric
moves according to the same equation as a charged particle moving radially in an
R--N spacetime. Consequently, the reversal of collapse to expansion cannot occur
between the two R--N horizons. Our example is consistent with this. Fig.
\ref{plothors} shows the radii of the outer horizon and of the maximal radius
achieved by a given ${\cal M}$ shell as functions of $x = N/N_0$. At the scale
of Fig. \ref{plothors}, the minimal radius and the radius of the inner horizon
seem to coincide with each other and with the $x$-axis. Fig. \ref{plothormin}
shows a closeup view of the two curves, and shows that indeed the inner horizon
has a smaller radius everywhere, except at the maximum, where the two curves are
tangent to each other. This is no numerical coincidence -- the two curves must
be tangent at every local extremum of $R_+$.

\begin{figure}
 \begin{center}
 \includegraphics[scale = 0.8]{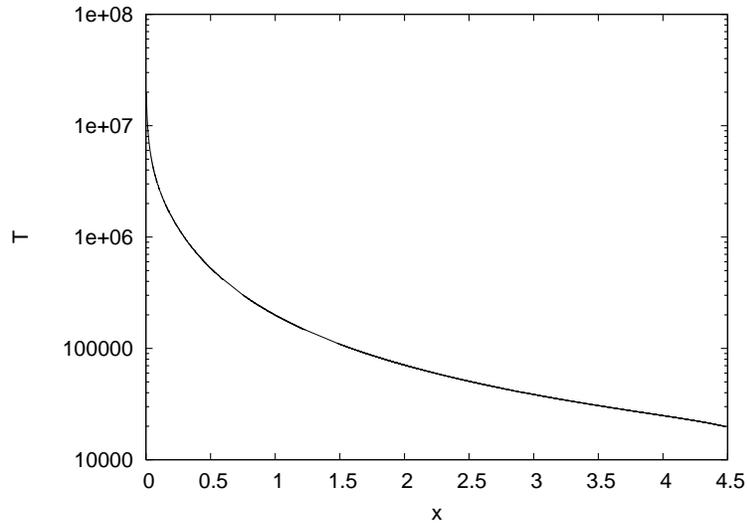}
 \caption{
 \label{plotperiod}
 \footnotesize
The period (in the time coordinate) as a function of $x = N/N_0$.}
 \end{center}
 \end{figure}

The limit of the period (in $s$) of oscillations in (\ref{8.8}) at $x \to 0$ is
infinite. Thus, the limiting period in the time coordinate will also be
infinite. Fig. \ref{plotperiod} shows the numerically calculated period in $t$
as a function of ${\cal M}$, in the range $x \in [0, 4)$ (for $x > 4$ the
calculation becomes progressively more difficult). As the figure shows, the
period decreases with increasing $x$ in this whole range, so we will assume $x =
4.0$ as a ``practical'' point of minimum of $T$. A real minimum must occur
somewhere at $x > 4.0$ because the period increases to $+ \infty$ at both $x \to
0$ and $x \to + \infty$.

\begin{figure}
 \begin{center}
 \includegraphics[scale = 0.8]{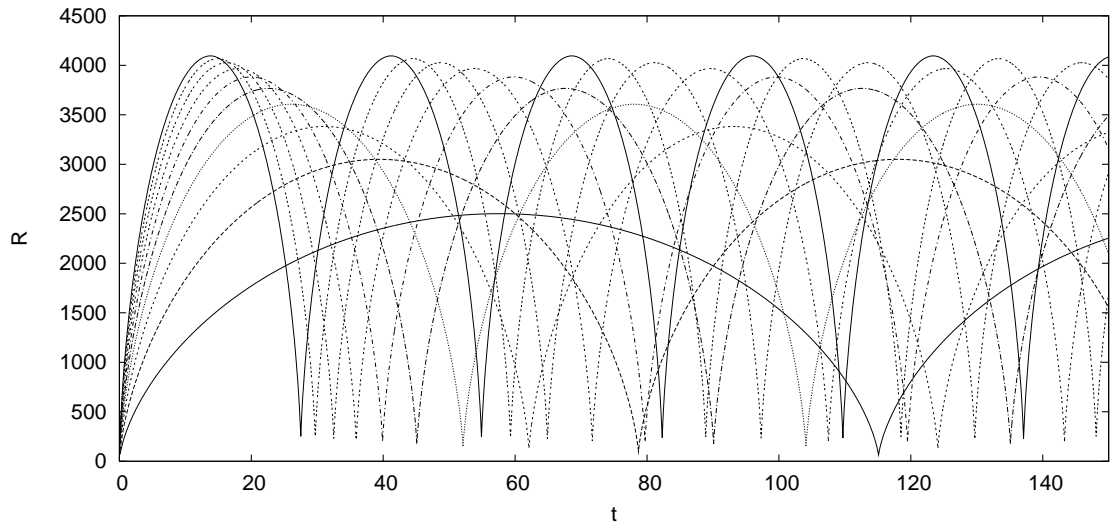}
 \caption{
 \label{drawcycles2}
 \footnotesize
The curves $R({\cal M}, t)$ corresponding to several values of ${\cal M}$. The
mass increases uniformly from $x = 0.01$ on the lowest curve to $x = 0.1$ on the
highest curve. The range of masses was chosen so as to make the figure readable;
the real range of periods and of amplitudes can be very large. See more
explanation in the text. The bounce is always smooth and at a nonzero value of
$R$. (The figure suggests otherwise, but this is only an illusion created by the
scale.) }
 \end{center}
 \end{figure}

Finally, Fig. \ref{drawcycles2} shows a collection of curves $R({\cal M}, t)$
corresponding to different values of ${\cal M}$, calculated numerically by
solving the set (\ref{2.21}) -- (\ref{2.23}). The configuration is
time-symmetric with respect to the instant $t = 0$. In agreement with Figs.
\ref{plothors}, \ref{plothormin} and \ref{plotperiod}, the maximal and minimal
radii achieved in each cycle are increasing functions of mass, while the period
is a decreasing function.\footnote{The range of masses in the figure is $x \in
[0.01, 0.1]$, where Figs. \ref{plothors}, \ref{plothormin} and \ref{plotperiod}
show that the radii and the period should really behave the way they do in Fig.
\ref{drawcycles2}. For Figs. \ref{plotperiod} and \ref{drawcycles2} the assumed
value of $b$ is 0.001.}

Figs. \ref{drawcycles2} and \ref{plotperiod} are drawn on the basis of numerical
solutions of the set (\ref{2.22}) -- (\ref{2.23}). As (\ref{2.22}) shows, for
small masses, where $Q \approx 0$, $C$ is nearly constant, i.e. nearly zero
(because, as mentioned earlier, $t$ can be chosen so that $C = 0$ at the
center). Thus, the dependence of $R$ on $t$ for small values of $x$ should be
very similar to the dependence of $R$ on $s$. Comparison of the corresponding
graphs confirms this: the graphs of $R({\cal M}, s)$ look identical to those of
Fig. \ref{drawcycles2} and the period as a function of $s$ has a graph that
looks identical to Fig. \ref{plotperiod}.

As predicted in Sec. \ref{nonbounce}, each mass shell avoids shell crossings
throughout the first expansion phase after the time-symmetric bounce, but then
experiences crossings after going through the maximal size. In order to avoid
shell crossings in the whole volume for the whole expansion phase of the
outermost shell, the radius of the dust ball cannot be too large. If it is very
large, then the period of oscillations of the outermost shells will also become
very large. Then, the time by which first shell crossings appear inside the ball
will become a small fraction of the duration of the expansion phase of the outer
surface, i.e. shell crossings will appear before the surface of the ball emerges
from inside the outer horizon.

\begin{figure}
 \begin{center}
 \includegraphics[scale = 0.7]{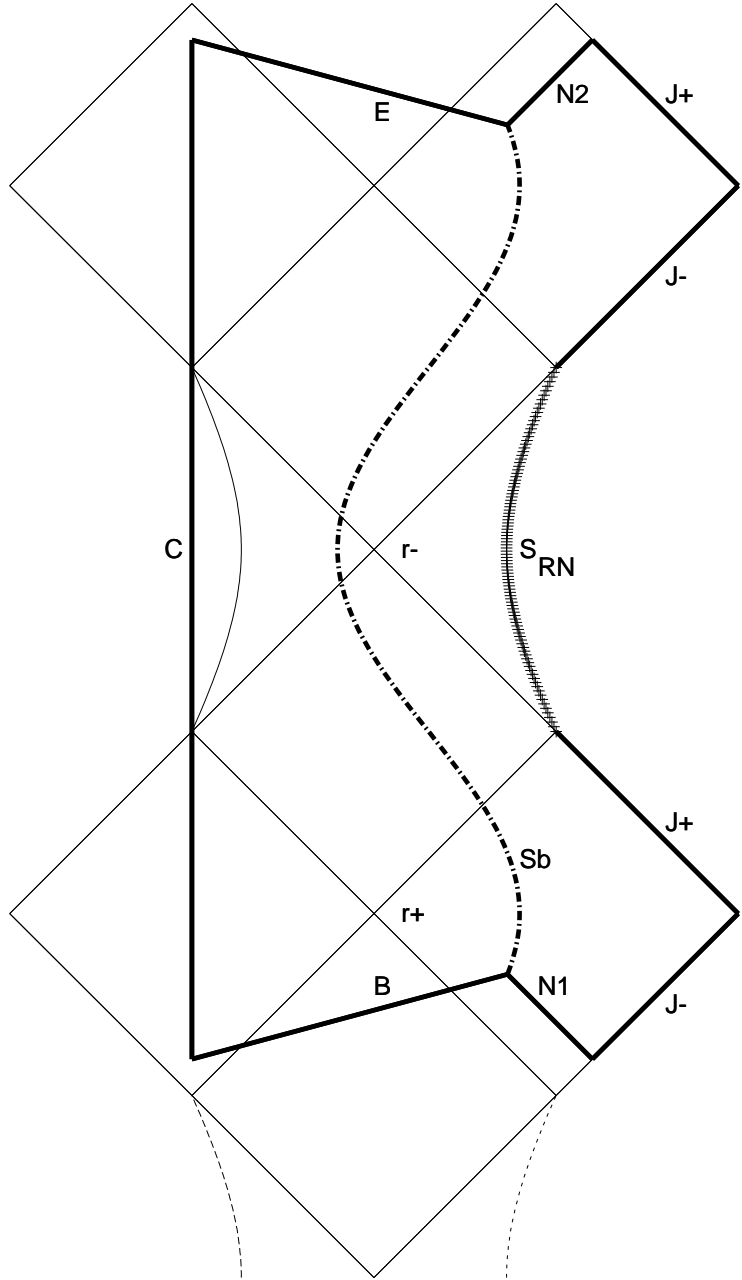}
 \caption{
 \label{chadumax}
 \footnotesize
A schematic Penrose diagram for the configuration defined by eqs. (\ref{11.1})
-- (\ref{11.2}) and (\ref{11.5}). See explanation in the text. }
 \end{center}
 \end{figure}

The evolution of our configuration is summarised in the Penrose diagram in Fig.
\ref{chadumax}. The diagram is written into the background of the Penrose
diagram for the maximally extended Reissner--Nordstr\"{o}m spacetime (thin
lines). C is the center of symmetry, Sb is the surface of the charged ball,
S$_{\rm RN}$ is the Reissner--Nordstr\"{o}m singularity. The interior of the
body is encompassed by the lines C, E, Sb and B; no singularity occurs within
this area. Lines B and E connect the points in spacetime where the shell
crossings occur at different mass shells. N1 (N2) are the past- (future-)
directed null geodesics emanating from the points in which the shell crossings
reach the surface of the body (compare Fig. \ref{drawcycles2}). The line Sb
should be identified with the uppermost curve in Fig. \ref{drawcycles2}. The top
end of Sb is where the corresponding curve in Fig. \ref{drawcycles2} first
crosses another curve, the middle point of Sb is at $t = 0$ in Fig.
\ref{drawcycles2}.

It would be interesting to have an example of a configuration that can pulsate
for ever, avoiding shell crossings in all of its collapse/expansion phases.
However, our example does not obey (\ref{10.4}) -- substitution of our functions
into (\ref{10.4}) leads to a clearcut contradiction. Thus, it is impossible to
impose on our example the condition that the period (in the proper time $s$) of
pusations is independent of ${\cal M}$. Whether such a permanently pulsating
singularity-free configuration exists at all is a problem to be investigated in
the future.

We recall that our example was only meant to demonstrate that the inequalities
in conditions (1) -- (9) are not mutually contradictory and allow a solution.
Since it turned out that the set of models is not empty, more examples should
exist, and now it is a challenge to explore other possibilities.

In the Friedmann limit (i.e. zero charge density, zero central charge and
homogeneous mass density), a model with $E < 0$ goes over to the $k > 0$
Friedmann model, for which the most natural topology of spatial sections is that
of $S^3$. Here we have a model that, if it were not matched to an R--N
spacetime, would have spatial sections of infinite volume (as attested by the
fact that it can contain an infinite amount of rest mass). Lema\^{\i}tre--Tolman
models with such spatial topology are known and understood geometrically
\cite{KrHe2004b}. However, the limiting transition from such a model to the $k >
0$ Friedmann model can be done only locally, i.e. in finite volumes, and
involves a discontinuity in the arbitrary functions. Thus, it would be desirable
to find a charged dust model with $E < 0$ that would have complete spatial
sections of finite volume. Curiously, it turned out to be much more difficult in
this case to fulfil conditions (7) -- (9), and no example of such a model has
been found. (The authors do not wish to imply that such an example does not
exist, this is simply a problem to be solved.)

\section{Can such an object exist in the real Universe?}

\setcounter{equation}{0}

There is a belief in the astronomical community that ordinary astronomical
objects, such as stars and galaxies, have zero net charge. Quantitative
estimates based on measurements and observations are, however, hard to come by.
The best that can be found in the literature are theoretical considerations on
how charges could be separated within the body of a star and what the maximal
charge could be. Before we compare our results with those predictions, we shall
first choose the most favourable parameters for our configuration.

It is clear that the smaller the net charge, the greater the chance that such an
object might exist. For our object, the absolute value of the charge first
increases from zero in the center to the maximum $\sqrt{G} N_0/(4c^2)$ achieved
at $x = 1$, and then keeps decreasing all the way to zero as $x \to \infty$. It
needs to be explained how such a charge could be concentrated toward the center.
We shall not dwell on this mechanism because stars are obviously not composed of
dust. In our case the concentration is a dynamical effect, achieved by
fine-tuning of the initial positions and velocities.

As seen from Fig. \ref{plotperiod}, the period of oscillations decreases with
mass, reaching a minimum at $x > 4.0$. Then, Fig. \ref{plothors} shows that at
$x = 4.0$, the maximal radius is slightly larger than the radius of the outer
horizon. The period in $s$, calculated from (\ref{8.8}), $2\pi M/(- 2E)^{3/2}$,
goes to $\infty$ both at $x \to 0$ and at $x \to \infty$, going through a
minimum at $x \approx 1.0$. The period in $t$ should behave similarly, i.e. will
start increasing at some $x = x_m$. We cannot take the radius of our object
larger than $x_m$ because if it is large, then the period will be large, and
shell crossings will appear inside the object before its surface emerges from
the outer horizon. This would be unsatisfactory -- we want the object to
distinctly emerge from the horizon before any singularities destroy it. Thus,
the largest total mass that we can assume corresponds to $x \approx 4.0$. At $x
= 4.0$, the charge is $|Q| = 0.16 \sqrt{G} N_0 / c^2$, while the active mass, in
physical units,\footnote{What we have been calling 'active mass' ${\cal M}$ is
in reality $(G/c^2)\times$ (mass).} is $c^2 {\cal M} / G = 1.68 N_0/c^2$. The
ratio of charge to mass is thus $G|Q|/(c^2 {\cal M}) = 0.095 \sqrt{G}$ in
electrostatic units (in which the unit of charge is $\sqrt{\rm g}$
cm$^{3/2}$/s). This makes $0.82 \times 10^{-14}$ coulombs per gram. For the
whole Earth, this ratio is $0.502 \times 10^{-25}$ C/g \cite{MiNo1995}. Compared
to the charge of the Earth, the charge of our object is thus enormous. However,
for a neutron star of 1 solar mass, the authors of Ref. \cite{RMLZ2004} found
that the total charge might be $10^{20}$ C, which makes $5.03 \times 10^{-14}$
C/g -- distinctly more than in our object. Thus, the possibility to find a real
object with charge and mass similar to our example is not
outlandish.\footnote{Note that we have not fixed the mass of our object -- the
constant $N_0$ is still arbitrary and can be made whatever we wish.} Only the
mechanism of charge separation remains a problem.

We stress again that the functions defining our object were chosen rather
accidentally, and lower ratios of charge to mass at the surface can possibly be
achieved. But the limiting ratio of charge to mass at the center must be
$\sqrt{G}$, which makes $8.616 \times 10^{-12}$ C/g.

The period of oscillation, found from (\ref{8.8}), is $T = 2 \pi M/(-
2E)^{3/2}$, but this is expressed in geometrical units. In physical units, the
period is $T/c$. Assuming that the surface of the object is at $x = 4.0$, taking
the corresponding mass to be ${\cal M} = 1.68 GN_0 / c^4$ and taking $b = 2.5$,
we calculate
 $$
T/c = 2.78 \times 10^{-38} \left(N_0/c^2\right),
 $$
which, for solar mass $N_0/c^2 = 1.989 \times 10^{33}$g, gives $5.52 \times
10^{-5}$s -- the same order of magnitude as the time of collapse to the
Schwarzschild singularity by a neutral star of one solar mass that has crossed
the horizon. Needless to recall, for an observer at infinity in the R--N
spacetime, the time it takes a collapsing object to reach its horizon is
infinite.

\section{Conclusions and possible further research}

We have shown that it is possible to set up such initial conditions for a
charged dust sphere of a finite radius that its outer surface completes one full
cycle of pulsation, while no singularity appears either at the surface or
anywhere inside it. The sequence of events is this:

1. The initial instant is when the outer surface of the sphere is close to the
maximum expansion. (When it is too late after the maximum, shell crossings in
the interior will appear before the full cycle is completed; when it is too
early before the maximum, shell crossings in the interior will exist already at
the initial instant). Other surfaces of constant mass reach their maxima of
expansion at (slightly) different instants of the time coordinate $t$. The outer
surface at the initial instant is outside the outer Reissner--Nordstr\"{o}m
horizon.

2. During collapse, the outer surface plunges first through the outer R--N
horizon, then through the inner R--N horizon, and bounces at a nonzero radius.
No shell crossings appear anywhere inside the sphere during the collapse phase,
and none of the constant-mass spheres collapse to zero radius, i.e. there are no
singularities in this phase.

3. For all constant-mass shells, the bounce at minimal size is simultaneous in
the time-coordinate t. This means that the evolution is time-symmetric with
respect to this instant. This makes the problem technically simpler: if there
were no singularities during collapse, there will be none during the next
expansion phase.

4. The expanding outer surface goes out of the inner R--N horizon, then out of
the outer R--N horizon and expands up to the maximal radius. As follows from the
analysis of the maximally extended R--N spacetime, after emerging from the outer
R--N horizon, the surface is in a different asymptotically flat sheet of the
extended R--N spacetime.

5. After the outer surface goes through the maximal radius, during the second
collapse phase, shell crossing singularities appear inside the ball and its
further evolution cannot be followed.

With the most favourable value of mass, the ratio of total charge to total mass
of our dust ball is only slightly larger than the corresponding theoretically
estimated maximum for charged neutron stars (see our Sec. 12 and Ref.
\cite{RMLZ2004}).

This is the description of our chosen example. However, other choices of the
arbitrary functions are possible that might improve some of the characteristics
of our model. For example, it would be desirable to avoid the matching to the
R--N spacetime, so that the dust distribution can extend over the whole space
(infinite or closed) -- then the solution could be investigated as a possible
model of the Universe with a localised charged object in it. Another desirable
generalisation would be to make {\it all} bounces time-symmetric, so that the
model oscillates singularity-free for ever. We shall now briefly discuss these
possibilities. Recall that one of the necessary conditions for the absence of
singularities is $E < 0$, and we shall assume this in the following.

Let us recall the necessary conditions for avoiding the BB/BC and shell-crossing
singularities. Two conditions are most important:

1. The absolute value of the ratio of charge density to mass density at the
center must be (in the units used here) $\sqrt{G}/c$ (see (\ref{2.20})).
Otherwise, either shell crossings are inevitable or the ratio of charge density
to mass density must be large throughout the volume to prevent BC.

2. The limit of the period of oscillations at zero radius must not be zero, so
that the pathological situation of Fig. \ref{cyclesfig} does not occur.

In addition to that, the following conditions must be obeyed:

3. The regularity conditions at the center of Sec. \ref{regcons}, which are the
required values of various functions at $N = 0$.

4. The 9 conditions listed in Sec. 10, most of which are inequalities to hold
throughout a neighbourhood of the center. Of these, the functions in conditions
(1) and (6 -- 9) must have zero limits at the center.

All these conditions are difficult to fulfil because there are only two
arbitrary functions in the model: ({\it i}) the charge distribution $Q(N)$, and
({\it ii}) the energy distribution $E(N)$ (which can equivalently be defined by
specifying $\Gamma(N)$ or ${\cal M}(N)$). Our choice here was the first $(Q, E)$
pair, found by trial and error, that obeys all the requirements, but only in a
limited volume and in a limited time interval.

To avoid the artificially limited volume, one might try to match the model to a
spacetime different from R--N. For example, it could be the Vickers model with
the charge density becoming strictly zero at a certain distance from the center.
Then, in the outer region, $Q,_N = 0$, which is sufficient to prevent the BB/BC
singularity. Since the main difficulty in avoiding shell crossings was close to
the center, and the $Q,_N = 0$ region would not extend to the center, chances
are that shell crossings could be avoided as well. In the extreme case, the
outer region could be the Lema\^{\i}tre--Tolman spacetime. Since the L--T
spacetime has strictly zero electric field, the function $Q(N)$ would have to be
zero at the surface of the charged region. The zero total charge would be an
advantage from the point of view of astronomy. We do not know if such a
configuration exists.

To avoid the limited time interval, the bounce at minimal radius should be
simultaneous in $t$ for all constant-mass shells. This means that the period of
oscillations (measured by the time-coordinate $t$) would have to be independent
of mass. Technically, this means that the integral of (\ref{9.1}) from the
minimal $R_+$ to the maximal $R_-$ (both given by (\ref{4.3})) should be
independent of ${\cal M}$. We have not been able to tackle this condition.

The condition to avoid zero limit of the oscillation period at the center is,
however, quite simple. It is reasonable to require that the function $C$ has a
finite value at the center, where $N = 0$ -- without this assumption, finite
intervals of the time coordinate $t$ would correspond to infinite intervals of
the proper time $s$. If this is the case, then it is enough to choose such
functions that the period in $s$ has a nonzero limit at $N = 0$. From
(\ref{8.8}), the condition for this is that $M / (- 2E)^{3/2}$ has a nonzero
limit at $N \to 0$. As noted at the very end of Sec. 10, the necessary condition
for this is: the limit calculated in (\ref{6.4}) is zero.

\section*{Acknowledgements}

We thank Amos Ori for extended correspondence on several subjects concerning
charged dust, in particular on the shell crossings and the existence of
nonsingular solutions. The discussion helped to clarify a few points that have
been left unexplained by previous publications.

\end{document}